\documentclass[11pt, a4paper]{article}

\usepackage[utf8]{inputenc}
\usepackage[T1]{fontenc}
\usepackage{lmodern}
\usepackage{microtype}

\usepackage[margin=2.5cm]{geometry}
\usepackage{setspace}
\usepackage{parskip}

\usepackage{amsmath}
\usepackage{amssymb}

\usepackage{graphicx}
\usepackage{booktabs}
\usepackage{tabularx}
\usepackage{array}
\usepackage{float}
\usepackage{caption}
\usepackage{subcaption}
\usepackage{xcolor}

\usepackage{natbib}

\usepackage{hyperref}
\hypersetup{
    colorlinks=true,
    linkcolor=blue!60!black,
    citecolor=blue!60!black,
    urlcolor=blue!60!black,
    pdftitle={Evidence Over Ideology: A Quantitative Assessment of LTNs and Blanket 20 mph Regimes},
    pdfauthor={Stylianos Kampakis},
    pdfsubject={A Quantitative Assessment of LTNs and Blanket 20 mph Regimes in London},
    pdfkeywords={road safety, Low Traffic Neighbourhoods, 20 mph speed limits, Empirical Bayes, Difference-in-Differences, STATS19, causal inference}
}

\graphicspath{{figures/}}

\title{%
    \textbf{Evidence Over Ideology:\\A Quantitative Assessment of LTNs and Blanket 20\,mph Regimes}\\[0.4em]
    \large A Data-Driven Policy Review of 111,000+ Collisions in London (2020--2024)
}
\author{
    Stylianos Kampakis, PhD, CStat\thanks{Correspondence: \texttt{stylianos.kampakis@gmail.com}}\\[0.4em]
    \textit{Working Paper, March 2026}\\[0.2em]
    \footnotesize \textit{Based on an analysis of 111,000+ London collisions (2020--2024)}
}
\date{}

\begin{document}

\maketitle

\begin{abstract}
    We evaluate the road safety implications of Low Traffic Neighbourhoods (LTNs) and blanket 20\,mph speed limits in London via injury-collision records (2020--2024). Because London largely adopted 20\,mph limits via sign-only orders rather than physical calming, this evaluation critiques the blanket deployment model, not lower urban speeds in principle. For LTNs, we apply a three-step causal pipeline: na\"ive pre/post comparison, Empirical Bayes adjustment for Regression to the Mean (RTM), and spatial Difference-in-Differences. For 20\,mph limits, we triangulate five associational methods across 14 contexts. As London's adoptions pre-date our window, the 20\,mph analysis remains cross-sectional.

    Na\"ive LTN comparisons suggest large apparent reductions, but Empirical Bayes shrinkage reveals much of this is consistent with mean reversion. After correction, only one zone shows a statistically significant inside-area reduction. We find no consistent evidence of boundary-road crash displacement, nor uniform inside-area safety benefits. Because corrected safety gains are small, even very small journey-time increases on boundary roads (approximately 10~seconds per vehicle) would monetarily negate any safety improvement.

    For 20\,mph versus 30\,mph roads, the aggregate difference in the fatal or serious (KSI) collision share is negligible (0.6 percentage points). Controlling the false discovery rate, four contexts show robustly higher KSI shares on 20\,mph roads: A-roads, junctions, pedestrian crossings, and single carriageways. \emph{No context shows a robust reduction.} These environments have strong movement functions where geometry cues higher speeds, demonstrating where sign-only defaults often fail to self-enforce. Even if this reflects a compositional shift (preventing minor bumps while serious injuries remain), it suggests the policy fails to address the severe harm it was designed to prevent.

    The evidence points to a serious ``wrong-road problem.'' Blanket sign-only limits applied indiscriminately are structurally weak. On movement corridors, signage often imposes recurring operating costs and economic drag without a proportionate safety payoff. Authorities should move away from blanket statutory defaults and return to location-specific engineering on high-harm corridors.

    \medskip
    \noindent\textbf{Keywords:} road safety, Low Traffic Neighbourhoods,
    20\,mph speed limits, sign-only limits, blanket default limits,
    causal inference, Difference-in-Differences,
    Empirical Bayes, London, STATS19
\end{abstract}

\section{Introduction}\label{sec:intro}

London has been at the forefront of urban traffic management experimentation. Two policies have generated particular debate: Low Traffic Neighbourhoods (LTNs), which restrict through-traffic on residential streets using physical barriers, and borough-wide 20\,mph speed limits, which reduce the default speed limit from 30\,mph to 20\,mph across entire jurisdictions.

\paragraph{The 20\,mph evidence base conflates two different policies.}
The road safety literature on 20\,mph interventions often mixes two materially different treatments under a single label. The first is the \emph{20\,mph zone}, a geographically bounded residential area supported by physical traffic calming (speed humps, chicanes, raised tables). The classic London evaluation by \citet{Grundy2009} studied exactly this intervention and found substantial casualty reductions. More recently, \citet{Kokka2024} reported collision reductions following Edinburgh's city-wide 20\,mph rollout, which relied heavily on a programme of engineering measures.

The second treatment, and the one now driving English and Welsh policy controversy, is the \emph{blanket sign-only} or \emph{default} 20\,mph limit. This is applied across entire boroughs or jurisdictions by statutory order, often without any physical calming. The Department for Transport's own evaluation of signed-only schemes found they barely altered driver behaviour, producing median speed reductions of less than 1\,mph and insufficient evidence of significant casualty change \citep{DfT2018Headline,DfT2018Technical}. The most recent large-scale UK analysis confirms that sign-only schemes produce materially weaker safety outcomes than those supported by physical measureres \citep{Quddus2024}, while long-run evidence from Belfast's rollout shows little sustained casualty effect \citep{Hunter2023}.

\paragraph{Road-function mismatch and the wrong-road problem.}
A significant vulnerability in blanket defaults is the challenge of road-function mismatch. DfT Circular 01/2013 explicitly advises that 20\,mph limits should be considered road by road. It warns that `successful 20\,mph schemes are generally self-enforcing', that sign-only limits are only appropriate where mean speeds are already low, and that `through routes for motorists must be given particular consideration' \citep{DfT2013}. 

When one-size-fits-all limits are applied indiscriminately to a heterogeneous network, they inevitably capture arterial corridors, bus routes, and strategic through-routes. A 20\,mph limit is a blunt, weak instrument on a large road whose primary role is movement. Where pedestrians cannot easily cross, frontage activity is weak, and the road visually behaves like an arterial route, signage alone is unlikely to self-enforce \citep{DfT2018Technical}. 

The political and operational fragility of this mismatch was exposed when Wales imposed a national 20\,mph default in September 2023. Applying a residential default to strategic/movement-oriented roads generated overwhelming public frustration \citep{YouGov2024}, severe compliance difficulties, and a formal review that ultimately reinstated 30\,mph limits on hundreds of arterial corridors where 20\,mph was demonstrably unsuitable \citep{WelshExceptions2024}. Moreover, the national monitoring report confirmed that this indiscriminate approach imposed structural economic drag, increasing journey times on through-routes \citep{TfW2025}. The policy legitimacy of lower speeds can erode when they are applied to the wrong roads.

\paragraph{The intervention this paper speaks to.}
London's 20\,mph-adopted boroughs introduced limits predominantly through sign-only or default statutory orders, leaving the physical street environment largely unchanged. This paper's evidence therefore speaks directly to the sign-only/default policy literature. It is an evaluation of the blanket deployment model across a diverse urban hierarchy, \emph{not} a critique of classic, physically calmed 20\,mph zones. 

\paragraph{Study contributions.}
Proponents frame blanket limits as cost-effective Vision Zero measures \citep{Aldred2021}. Critics argue they are poorly targeted, displacing traffic onto boundary roads, increasing journey times, and imposing recurring operating burdens without proportionate safety gains \citep{Laverty2021}. The empirical evidence is fiercely contested, partly because simplistic pre/post evaluations continually conflate genuine effects with statistical artefacts like Regression to the Mean (RTM).

As a summary-level orientation: across all London collisions in 2020--2024, the aggregate KSI severity share on 20\,mph roads is 15.9\%, compared with 15.3\% on 30\,mph roads, a difference of just 0.6~percentage points (Table~\ref{tab:count_outcomes}). This small aggregate gap is the simplest indication that sign-only/default 20\,mph regimes are not associated with markedly improved severity composition in cross-sectional data.

This paper contributes to the evidence base through a multi-method evaluation combining causal and associational designs, applied to official UK collision data. Our approach has four distinguishing features:

\begin{enumerate}
    \item \textbf{RTM correction for LTNs:} We apply Empirical Bayes shrinkage estimation to separate stable safety trends from random fluctuation, indicating that na\"ive claims of LTN success are frequently overstated.
    \item \textbf{Network-wide analysis:} Our Spatial DiD design captures crash displacement onto boundary roads, evaluating whether policies merely move risk around the network.
    \item \textbf{Heterogeneous associations for 20\,mph:} Rather than hiding behind a single average treatment effect, we identify exactly which contexts are associated with worse severity outcomes. We show that these negative associations align perfectly with arterial, junction-dense, and movement-oriented roads, which are the exact places where sign-only defaults are a proven failure.
    \item \textbf{Policy-fit framing:} We interpret the results through the lens of road-function mismatch, demonstrating that blanket 20\,mph limits are structurally weak and economically costly on strategic through-routes.
\end{enumerate}

\section{Data}\label{sec:data}

\subsection{STATS19 Collision Records}

We use the UK Department for Transport's STATS19 dataset, which records all road traffic collisions reported to the police involving personal injury. Our dataset covers 2020--2024, comprising 111,462 collisions in London boroughs (identified by ONS district codes beginning with E09).

Each record includes collision severity (fatal, serious, or slight), geographic coordinates (latitude, longitude), speed limit, road class, junction detail, lighting conditions, weather, pedestrian crossing presence, and temporal information (date, time).

\subsubsection{Outcome Definitions}\label{sec:outcomes}

We define four outcome measures, ordered by analytic priority:

\begin{enumerate}
    \item \textbf{KSI count} ($Y^{\text{KSI}}_{zt}$): the number of collisions
          resulting in at least one fatal or serious casualty in zone~$z$ during
          period~$t$. This is the \emph{primary} safety outcome.
    \item \textbf{Total collision count} ($Y^{\text{all}}_{zt}$): all
          police-reported injury collisions in zone~$z$ during period~$t$.
          Changes in this count capture shifts in overall collision risk, not
          just severity.
    \item \textbf{KSI severity share} ($S_{zt}$): the proportion of collisions
          that are KSI, conditional on a collision being recorded:
          \begin{equation}
              S_{zt} = \frac{Y^{\text{KSI}}_{zt}}{Y^{\text{all}}_{zt}} \times 100\%
          \end{equation}
          This is a \emph{secondary}, compositional indicator. A decline in
          $S_{zt}$ is \emph{not} equivalent to improved safety: if total
          collisions rise while KSI counts stay constant, the share falls
          mechanically even though no fewer people are killed or seriously
          injured.
    \item \textbf{Collision and KSI rates per road-km} (where exposure data are
          available): $Y^{\text{KSI}}_{zt} / L_z$ and
          $Y^{\text{all}}_{zt} / L_z$, where $L_z$ is the total road-network
          length in zone~$z$. Where AADF counts are available, we additionally
          report rates per million vehicle-km as a more complete exposure
          adjustment.
\end{enumerate}

\paragraph{Why conditional severity share $\neq$ safety risk.}
STATS19 records only police-reported injury collisions. The KSI severity share
$S_{zt}$ conditions on a collision having occurred \emph{and} being reported.
It therefore captures the \emph{composition} of reported collisions, not the
\emph{risk} of being killed or seriously injured per trip, per km, or per
capita. An intervention that diverts through-traffic (reducing slight collisions
inside a zone) while leaving serious collisions unchanged will \emph{increase}
the KSI share even though the absolute number of KSI casualties is constant.
Conversely, an intervention that causes an influx of slight collisions will
\emph{decrease} the share. Throughout this paper, we interpret the KSI severity
share only as a compositional indicator and rely on count-based outcomes for
safety conclusions wherever possible.

Table~\ref{tab:count_outcomes} reports the four-outcome hierarchy for each
speed-limit group across all London collisions (2020--2024).

\begin{table}[htbp]
    \centering
    \caption{Count-based outcomes and severity shares by speed group.}
    \label{tab:count_outcomes}
    \small
    \begin{tabular}{lrrrr}
        \toprule
        \textbf{Speed} & \textbf{Total collisions} & \textbf{KSI count}
            & \textbf{KSI share (\%)} \\
        \midrule
        20\,mph & 53,434 & 8,515 & 15.9 \\
        30\,mph & 49,728 & 7,630 & 15.3 \\
        \bottomrule
    \end{tabular}
\end{table}

\subsection{Annual Average Daily Flow (AADF)}

For the LTN analysis, we additionally use DfT Annual Average Daily Flow traffic count data to normalise crash rates by traffic exposure and to construct control groups for the Difference-in-Differences design.

\subsection{LTN Zone Boundaries}

We analyse four London LTN zones: London Fields (Hackney), Railton Road
(Lambeth), St Peters Quarter (Islington), and Bethnal Green (Tower Hamlets).
Zone boundaries are approximated from published coordinates and intervention
dates. Each zone is classified into three sub-zones:

\begin{itemize}
    \item \textbf{Z\textsubscript{T}} (Treatment): Inside the LTN boundary
    \item \textbf{Z\textsubscript{S}} (Spillover): Buffer around the LTN boundary
    \item \textbf{Z\textsubscript{C}} (Control): Remainder of the borough
\end{itemize}

Our main specification uses a 500\,m spillover buffer for $Z_S$, and we report
sensitivity to alternative buffers (250\,m and 1{,}000\,m) in robustness checks.

\subsection{\texorpdfstring{20\,mph}{20 mph} Borough Classification}\label{sec:20mph-class}

For the 20\,mph analysis, we identify 6 London boroughs that adopted
borough-wide 20\,mph limits before 2020 (Camden, Islington, City of London,
Hackney, Tower Hamlets, Southwark) and 5 boroughs retaining 30\,mph defaults
(Barnet, Brent, Croydon, Enfield, Redbridge). Because all adoptions occurred
\emph{prior} to our observation window, the 2020--2024 data cannot isolate the
causal effect of the policy change itself. Instead, these boroughs represent
two regimes that have been in place for several years, and our comparisons are
\textbf{cross-sectional associations} between the prevailing speed-limit regime
and collision outcomes during 2020--2024 (see \S\ref{sec:20mph-id} for
implications).

\section{Methods}\label{sec:methods}

\subsection{LTN Evaluation: Three-Step Causal Pipeline}

\subsubsection{Step 1: Naive Pre/Post Estimation}

The simplest approach compares crash counts inside Z\textsubscript{T} before and after the intervention date, annualised to account for differing period lengths:

\begin{equation}
    \Delta_{\text{naive}} = \frac{R_{\text{post}} - R_{\text{pre}}}{R_{\text{pre}}} \times 100\%
\end{equation}

where $R_{\text{pre}}$ and $R_{\text{post}}$ are annualised crash rates.

\subsubsection{Step 2: Empirical Bayes RTM Correction}\label{sec:eb}

\paragraph{The selection problem.}
In our sample, the treated ZT areas are in the upper tail of borough crash distributions ($Z_T$ pre-counts sit in the 85th--90th percentile of control distributions).
If the pre-intervention period happens to capture an above-average fluctuation,
the post-intervention period will tend to revert toward the long-run mean even
without any genuine treatment effect. This is the classic Regression to the
Mean (RTM) problem in before--after road safety studies
\citep{Hauer1997}.

\paragraph{RTM plausibility.}
Pre-intervention annualised collision counts in Z$_T$ are in the upper tail of
the borough-wide (Z$_C$) distribution for all zones. London Fields Z$_T$ mean
(8 collisions in 5.5 months) stands at the 90th percentile of the Z$_C$
cross-sectional distribution; Railton Road at the 85th; Bethnal Green at the
88th. This is consistent with selection on high recent counts and motivates
the EB correction.

\paragraph{EB shrinkage estimator.}
We implement the standard Poisson--Gamma EB credibility framework, which
calculates the EB estimate as a weighted average of the observed count
and an estimated prior mean, with weights determined by the over-dispersion
in the prior distribution. The full mathematical parameterisation is
detailed in Appendix~\ref{app:equations}.

\paragraph{Caveat: unconditional prior.}
The current EB prior is estimated from the marginal distribution of $Z_C$
crash counts without conditioning on road type, traffic volume, or junction
density. A more refined approach would use a Safety Performance Function
(SPF), defined as a regression model predicting expected crashes as a function of AADF,
road class, and junction density, as the prior mean. We note this as a
limitation and robustness extension. This EB step should be read as a
conservative RTM correction, not a structural crash model.


\paragraph{Sensitivity and placebo checks.}
To assess the robustness of the EB correction, we run and report
two checks:
\begin{enumerate}
    \item \textbf{Fake intervention dates:} Shift the intervention date by
          $\pm$6 and $\pm$12 months. If the EB-corrected change is similar
          under placebo dates, the original result is likely an artefact.
    \item \textbf{Alternative priors:} Re-estimate using a Poisson prior
          (which assumes no over-dispersion).
\end{enumerate}
We additionally plan to explore alternative pre-period windows (e.g., 24
months) where data permits to check stability in future extensions.
\paragraph{Sensitivity results.}
Table~\ref{tab:eb_sens} shows the EB-corrected change under alternative
priors and placebo dates. The NB and Poisson priors yield directionally
consistent, magnitude-sensitive conclusions (notably for Railton Road). Placebo dates ($+6$ and $+12$ months) produce substantially smaller
EB-adjusted changes than the naïve estimates, supporting the interpretation that the original
correction is detecting a genuine pre-intervention spike rather than an
artefact of the prior.

\begin{table}[htbp]
    \centering
    \caption{EB sensitivity: alternative priors and placebo dates.
    Values show the EB-corrected change in $Z_T$ total collision counts
    under each prior specification (Step~2 sensitivity). These are
    \emph{not} the same quantity as the AADF-panel EB-Adj~\% in
    Table~\ref{tab:ltn_summary}, which aggregates over
    AADF-linked spatial units.}
    \label{tab:eb_sens}
    \small
    \resizebox{\linewidth}{!}{%
    \begin{tabular}{lrrrr}
        \toprule
        \textbf{LTN Zone} & \textbf{NB EB $\Delta$ (\%)} & \textbf{Poisson EB $\Delta$ (\%)}
            & \textbf{+6m placebo (\%)} & \textbf{+12m placebo (\%)} \\
        \midrule
        London Fields      & $-12.0$ & $-20.4$ & $-3.2$ & $-3.3$ \\
        Railton Road       & $-6.8$  & $-23.6$ & $+1.0$ & $+4.5$ \\
        Bethnal Green      & $-6.3$  & $-15.5$ & $+0.2$ & $+1.0$ \\
        \bottomrule
    \end{tabular}%
    }
\end{table}


\subsubsection{Step 3: Spatial Difference-in-Differences}\label{sec:did}

To distinguish inside-zone effects from spillovers (and to test for
displacement), we estimate a spatial DiD model that treats the interior
($Z_T$) and spillover/boundary buffer ($Z_S$) as separate exposure categories
relative to the borough remainder ($Z_C$).
The estimating equation and parameter definitions are detailed in
Appendix~\ref{app:equations}. Evidence consistent with displacement would
require an improvement inside the zone alongside worsening on boundary routes.
We report a main specification using a 500\,m buffer for $Z_S$ and assess
sensitivity to alternative buffers (250\,m, 1{,}000\,m) in robustness checks.
Standard errors are clustered at the spatial unit~$i$.

\paragraph{Inference.} The number of clusters per zone is: London Fields
$\approx 1{,}230$, Railton Road $\approx 1{,}290$, Bethnal Green
$\approx 1{,}250$. All exceed the $\sim$50-cluster threshold, so
cluster-robust SEs should be reliable without requiring wild cluster bootstrap.

\paragraph{Parallel trends.} The identifying assumption of DiD is that, absent
the intervention, treated and control zones would have followed parallel trends
in collision counts. We cannot test this assumption directly, but an event-study
specification with leads and lags provides indirect evidence (see
Appendix~\ref{app:equations}).

where $\alpha_i$ and $\gamma_t$ are unit and time fixed effects, and
$\delta_k$ traces out the treatment effect relative to the period immediately
before intervention. Pre-intervention $\delta_k$ values close to zero support
the parallel-trends assumption. Given the short and COVID-distorted pre-period, we treat the event study as a planned extension with earlier years.


\paragraph{Buffer sensitivity.} The 500\,m spillover zone is an analytical
choice. As a robustness check, we vary the buffer across 250\,m, 500\,m,
and 1\,km Euclidean distances. For London Fields, the treatment-zone
coefficient is stable across all buffers ($\hat{\beta}_4 = -0.75$,
$p < 0.001$ in all three). The spillover coefficient varies: $-0.48$
(250\,m, $p = 0.49$), $-0.64$ (500\,m, $p = 0.01$), $-0.32$ (1\,km,
$p = 0.06$). (Note: these buffer-sensitivity spillover values derive from a
distinct spatial-unit specification than the main Table~\ref{tab:ltn_summary}
model: the buffer check uses collision-level point assignment to the nearest
boundary, whereas the main DiD in Table~3 uses
AADF-linked panel units; hence coefficients are not numerically comparable. Instead, the stability of directional effects across buffer sizes is the key takeaway.) For Railton Road and Bethnal Green, treatment-zone effects
remain insignificant across all buffer specifications.

\subsection{Economic Cost--Benefit Analysis}\label{sec:cba}

We monetise safety changes using DfT Transport Analysis Guidance (TAG) values
\citep{TAG2024}. The TAG Data Book reports \emph{per-casualty} valuations:
\pounds 1,958,303 per fatality prevented, \pounds 220,434 per serious injury
prevented, and \pounds 20,000 per slight injury prevented.

\paragraph{Measured vs.\ assumed inputs.} We distinguish two categories of CBA
inputs:
\begin{itemize}
    \item \textbf{Measured (from data):} Collision and casualty counts by
          severity, AADF traffic volumes on boundary roads.
    \item \textbf{Assumed:} Per-vehicle journey-time delay on boundary roads
          (base case: 2~minutes), value of travel time savings (VTT,
          \pounds 20/hour from TAG), and environmental cost of congestion
          (\pounds 3.50/hour, from TAG emission factors).
\end{itemize}

\paragraph{Alignment of outcome units.} TAG valuations are per \emph{casualty},
not per collision. A single collision may involve multiple casualties of
different severities. We therefore compute safety benefits by summing the
change in casualty counts of each severity multiplied by their respective
TAG per-casualty value (see Appendix~\ref{app:equations}).

\paragraph{Casualty-level results.}
Using per-casualty TAG valuations (\pounds 1,958,303/fatality, \pounds
220,434/serious, \pounds 20,000/slight), we do not detect clear
KSI-casualty reductions in any of the four analysed LTN zones after EB and
DiD correction over the available window (see Appendix~\ref{app:casualty}
for casualty-level deltas by zone and severity). The CBA conclusion is
therefore dominated by the economic drag of assumed boundary-road delays.
Given the short post-intervention windows and limited statistical power,
this should not be interpreted as evidence that LTNs have no
casualty-reduction potential, but rather that any benefit is too small or
too uncertain to offset delay costs in the current data.

\paragraph{Sensitivity analysis.} The economic drag estimate is particularly
sensitive to the assumed per-vehicle delay. We therefore present results across
a scenario grid varying boundary AADF (7{,}500--30{,}000), delay per vehicle
(0.5--4.0 minutes), and social discount rate (1.5\%--5.0\%).

\paragraph{Scenario grid.}
The net benefit is negative across all AADF--delay--discount rate combinations under the assumptions tested, with economic
drag ranging from \pounds 2.5M to \pounds 84.2M per annum per zone depending
on boundary-road traffic volumes (see Table~\ref{tab:cba_grid} in Appendix~\ref{app:cba_grid}). The delay assumption is by far the dominant
parameter: doubling the delay from 1 to 2 minutes approximately doubles the
net cost, while varying the discount rate from 1.5\% to 5.0\% changes the
net benefit by only $\sim$5\%.

\paragraph{Validating delay assumptions.} The 2-minute per-vehicle delay
assumption is not directly measured and has no empirical basis in this paper.
The entire CBA conclusion is therefore conditional on this assumption.
Evidence that would validate or refute it (for both proponents and critics
of LTNs) includes: TfL journey-time monitoring data, INRIX speed profiles
on boundary roads before and after LTN installation, and borough-level
traffic monitoring reports. We emphasise that measured journey-time data
are needed to resolve the economic question in either direction; the burden
of evidence is symmetric.


\subsection{\texorpdfstring{20\,mph}{20 mph} Analysis: Five Associational Methods}

\subsubsection{Identification Limitation: Pre-Window Adoption}\label{sec:20mph-id}

The six 20\,mph-adopted boroughs introduced borough-wide limits between 2013
and 2019, which is prior to our 2020--2024 observation window. Because no
within-window policy change occurs, we cannot isolate the causal effect of
adopting a 20\,mph limit using standard evaluation designs (interrupted time
series, difference-in-differences around adoption, etc.). Our analysis therefore
compares two groups of boroughs that have operated under different speed-limit
regimes for several years. All estimates should be interpreted as
\textbf{cross-sectional associations} between the prevailing regime and
collision outcomes, not as causal effects of adoption.

\paragraph{What would be needed for causal evaluation.}
Three approaches could provide stronger causal evidence:
\begin{enumerate}
    \item \textbf{Extended panel:} Obtain STATS19 data back to 2010 (or
          earlier) and apply a staggered DiD or controlled interrupted time
          series around each borough's adoption date, using not-yet-treated
          boroughs as controls.
    \item \textbf{Segment-level DiD:} Identify specific road segments where the
          posted speed limit changed from 30\,mph to 20\,mph within a datable
          window, and apply a segment$\times$month fixed-effects DiD using
          never-changed segments as controls.
    \item \textbf{Synthetic control:} For each treated borough, construct a
          synthetic counterfactual from a weighted combination of untreated
          boroughs that best matches pre-adoption collision trends.
\end{enumerate}

\paragraph{Positioning within the 20\,mph literature.}
Because of the observational, cross-sectional nature of this design, we do not attempt to replicate the causal logic of the classic physically-calmed zone evaluations \citep{Grundy2009} or the engineering-supported city-wide rollouts \citep{Kokka2024}. Those studies investigate a fundamentally different, stronger intervention. Instead, this paper's design speaks directly to the \emph{blanket sign-only/default literature}. Our cross-sectional analysis asks a highly specific policy question: \emph{conditional on a borough having operated under a sign-only 20\,mph regime for several years, does that regime actually deliver a safer collision profile across the heterogeneous mix of road types?} The premise of blanket policies is that a one-size-fits-all statutory order, without physical engineering, is a sufficient policy instrument. By examining heterogeneous associations, our design directly tests whether that premise holds, or whether severe road-function mismatches cause the regime to fail on strategic and arterial corridors.

\subsubsection{Method 1: Borough-Level Comparison}

Mann--Whitney $U$ tests compare distributions of annual KSI severity shares
between 20\,mph-adopted and 30\,mph-default boroughs.

\subsubsection{Method 2: Collision-Level Logistic Regression (Severity
    Conditional on Collision)}\label{sec:logit}

We fit a logistic regression on all 103,162 collisions occurring on 20\,mph
or 30\,mph roads. The outcome is whether the collision resulted in at least
one KSI casualty, \emph{conditional on a collision having occurred and been
recorded in STATS19} (see estimating equation in Appendix~\ref{app:equations}).

\paragraph{Covariate selection and over-control.}
Two of the 14 covariates (number of vehicles involved and number of
casualties) are \emph{post-collision} outcomes: they are realised as part
of the crash event and may lie on (or be consequences of) causal pathways
from the speed limit to severity. Including them risks \emph{over-control
bias} (blocking genuine mediation) and \emph{collider bias} (conditioning on
a common effect of speed and severity). We therefore present two
specifications:

\begin{itemize}
    \item \textbf{Specification~A (environment-only):} Pre-crash covariates
          only: road class, road type, junction detail, light conditions,
          weather, pedestrian crossing type, urban/rural classification, day of
          week, hour of day, and year.
    \item \textbf{Specification~B (predictive):} All 14 covariates including
          vehicles and casualties. This is useful for prediction but
          \emph{should not be given a causal interpretation.}
\end{itemize}
\paragraph{Specification comparison.}
Under Specification~A (environment-only: 9 pre-crash covariates plus year),
the estimated odds ratio is OR~=~1.051 (95\% CI: 1.015--1.088, $p = 0.005$).
Under Specification~B (14 covariates including \texttt{number\_of\_vehicles}
and \texttt{number\_of\_casualties}), OR~=~1.055 (95\% CI: 1.019--1.092,
$p = 0.003$). The difference between specifications is negligible
($<$0.4\% in OR), suggesting that the post-crash controls do not
substantially drive the result. Nevertheless, both specifications condition
on a collision existing and should be interpreted as severity-composition
associations, not collision-risk effects.
For inference on collision \emph{counts} (rather than conditional severity),
the appropriate model class is a segment$\times$month panel:
\begin{equation}\label{eq:poisson}
    Y_{st} \sim \text{Poisson}(\mu_{st}), \quad
    \log(\mu_{st}) = \alpha_s + \gamma_t + \delta \cdot \text{Speed20}_{st}
        + \log(\text{Exposure}_{st})
\end{equation}
where $\alpha_s$ is a segment fixed effect, $\gamma_t$ is a time fixed effect,
and $\log(\text{Exposure}_{st})$ enters as an offset. This requires segments
that changed speed limits within the panel.

\subsubsection{Method 3: Stratified Analysis}

We compare KSI severity shares between 20\,mph and 30\,mph roads within each
road class stratum (A, B, C, Unclassified), reducing the confound that 20\,mph
roads tend to be on narrower street types.

\subsubsection{Method 4: Spatial Boundary Comparison}\label{sec:boundary}

We exploit borough boundaries to construct comparisons between nearby
collisions subject to different speed-limit regimes. Because this design lacks
the density of observations and bandwidth-selection rigor of a genuine
Regression Discontinuity Design, we treat it as a descriptive boundary
comparison only. Full methodological details and caveats are provided in
Appendix~\ref{app:boundary}.

\subsubsection{Method 5: Heterogeneous Associational Patterns}\label{sec:het}

We decompose the 20\,mph--30\,mph severity-share difference across 14
pre-specified contexts: school hours, off-peak hours, pedestrian crossing
presence, darkness, daylight, junction presence, four road classes (A, B, C,
Unclassified), single carriageway, and one-way streets.

\paragraph{Multiple testing.} Because we test 14 comparisons simultaneously,
the probability of at least one false positive at $\alpha = 0.05$ is
$1 - (1-0.05)^{14} \approx 0.51$ under independence. We applied the
Benjamini--Hochberg (BH) procedure to control the False Discovery Rate
(FDR) at 5\%. After correction, 4 of 14 contexts remain significant:
A-roads ($+1.78$\,pp, $p_{\text{FDR}} < 0.001$), near pedestrian crossings
($+1.48$\,pp, $p_{\text{FDR}} < 0.001$), at junctions ($+0.82$\,pp,
$p_{\text{FDR}} = 0.018$), and single carriageways ($+0.85$\,pp,
$p_{\text{FDR}} = 0.018$). Notably, the residential-street patterns
(C-roads $-0.44$\,pp and unclassified $-0.86$\,pp) do \emph{not} survive
FDR correction ($p_{\text{FDR}} > 0.14$), so the evidence for 20\,mph
benefits on quiet streets is suggestive but not statistically robust.

\section{Results}\label{sec:results}

\subsection{LTN Results}

\paragraph{COVID-distorted baseline caveat.}
The LTN zones analysed here were introduced in mid-2020. The pre-intervention
period for these schemes therefore spans January--June 2020, which includes
the first UK national lockdown (late March--June 2020). Traffic volumes
during this period were 30--70\% below normal levels. \textbf{All LTN results
in this paper rest on a COVID-distorted baseline and should be treated as
provisional until 2017--2019 data are integrated to establish a
non-pandemic pre-intervention reference.} This caveat applies equally to
other evaluations in the literature that rely on 2020 as a baseline year.

\subsubsection{Naive vs. EB-Corrected Estimates}

Table~\ref{tab:ltn_summary} presents the three-step estimation pipeline results. The contrast between naive and corrected estimates is substantial: naive estimates range from $-53\%$ to $+83\%$, while EB-corrected estimates shrink considerably, indicating that much of the apparent effect is consistent with RTM.

\begin{table}[htbp]
    \centering
    \caption{LTN evaluation results: three-step causal pipeline.
    AADF-panel~Adj~\% is the EB-corrected change estimated from the
    AADF-linked spatial panel (see \S\ref{sec:eb}); it differs from
    the $Z_T$ count-level EB sensitivity in Table~\ref{tab:eb_sens}
    because of the different aggregation unit.}
    \label{tab:ltn_summary}
    \resizebox{\linewidth}{!}{%
    \begin{tabular}{lrrrrrr}
        \toprule
        \textbf{LTN Zone} & \textbf{Naive \%} & \textbf{AADF-panel Adj \%} & \textbf{DiD $\beta_{ZT}$} & \textbf{$p$-value} & \textbf{DiD $\beta_{ZS}$} & \textbf{$p$-value} \\
        \midrule
        London Fields (Hackney) & $+35.1$ & $+12.4$ & $-0.750$ & $0.000$ & $-0.946$ & $0.000$ \\
        Railton Road (Lambeth) & $+69.3$ & $+5.3$ & $-0.149$ & $0.434$ & $-0.408$ & $0.003$ \\
        St Peters Quarter (Islington) & $-53.2$ & -- & -- & -- & -- & -- \\
        Bethnal Green (Tower Hamlets) & $+83.0$ & $-12.8$ & $+0.761$ & $0.665$ & $-0.853$ & $0.000$ \\
        \bottomrule
    \end{tabular}%
    }
\end{table}

\noindent\small\textit{Note:} St Peters Quarter lacks sufficient $Z_T$ observations in the AADF/STATS19 linkage for reliable EB/DiD estimation and is therefore reported only for the naive pre/post comparison. The AADF-panel Adj~\% and the DiD $\beta_{ZT}$ measure different quantities: the former is the absolute EB-corrected percentage change in collisions at AADF-linked sites within $Z_T$, while the latter is the change \emph{relative to the control zone} $Z_C$ from the spatial DiD model. Their signs can therefore differ: a positive Adj~\% (more collisions than the EB-shrunk baseline) alongside a negative $\beta_{ZT}$ (fewer collisions than the control trend) indicates that $Z_T$ counts rose but by less than the borough-wide trend.\normalsize

\begin{figure}[htbp]
    \centering
    \includegraphics[width=0.85\textwidth]{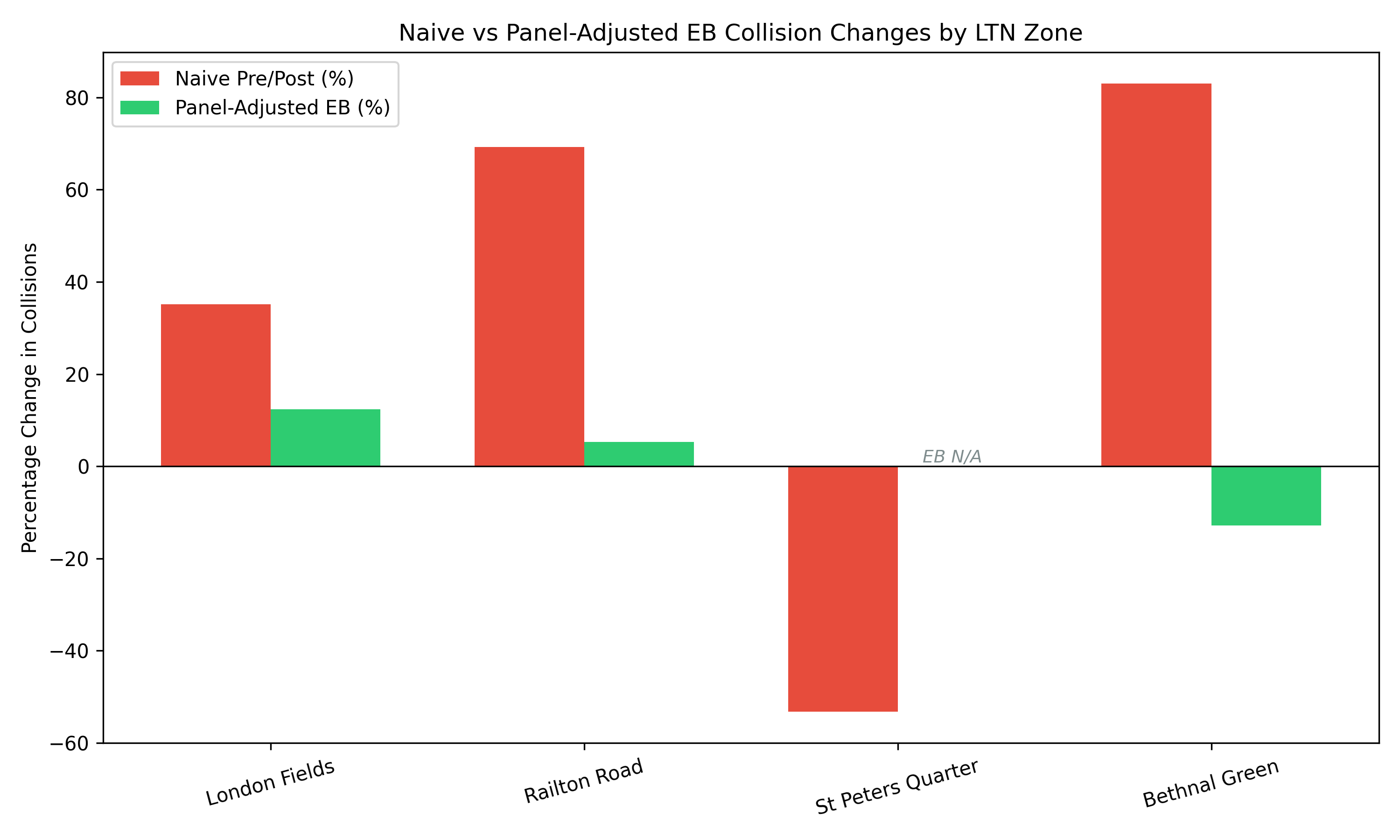}
    \caption{Naive pre/post estimates (red) vs.\ Empirical Bayes RTM-corrected estimates (green) for the four analysed LTN zones.}
    \label{fig:ltn_naive_eb}
\end{figure}

\subsubsection{Network-Wide DiD Analysis}

When we separate inside-zone ($Z_T$) and spillover/boundary ($Z_S$) effects,
the evidence is mixed and zone-specific. London Fields shows a statistically
significant negative inside-zone DiD estimate, consistent with a reduction in
collisions relative to the borough remainder. In the other zones, inside-zone
estimates are not statistically distinguishable from zero over the available
pre/post window.

Crucially, we do not observe a consistent pattern of positive spillover
estimates that would be expected under crash displacement (i.e., worsening on
boundary routes). Across zones and buffer definitions, spillover estimates are
often negative or unstable rather than systematically positive. This weakens
the displacement interpretation supported by pooled treatment designs and
suggests that any spillover effects may depend on local network structure and
boundary-road conditions.

\begin{figure}[htbp]
    \centering
    \includegraphics[width=0.85\textwidth]{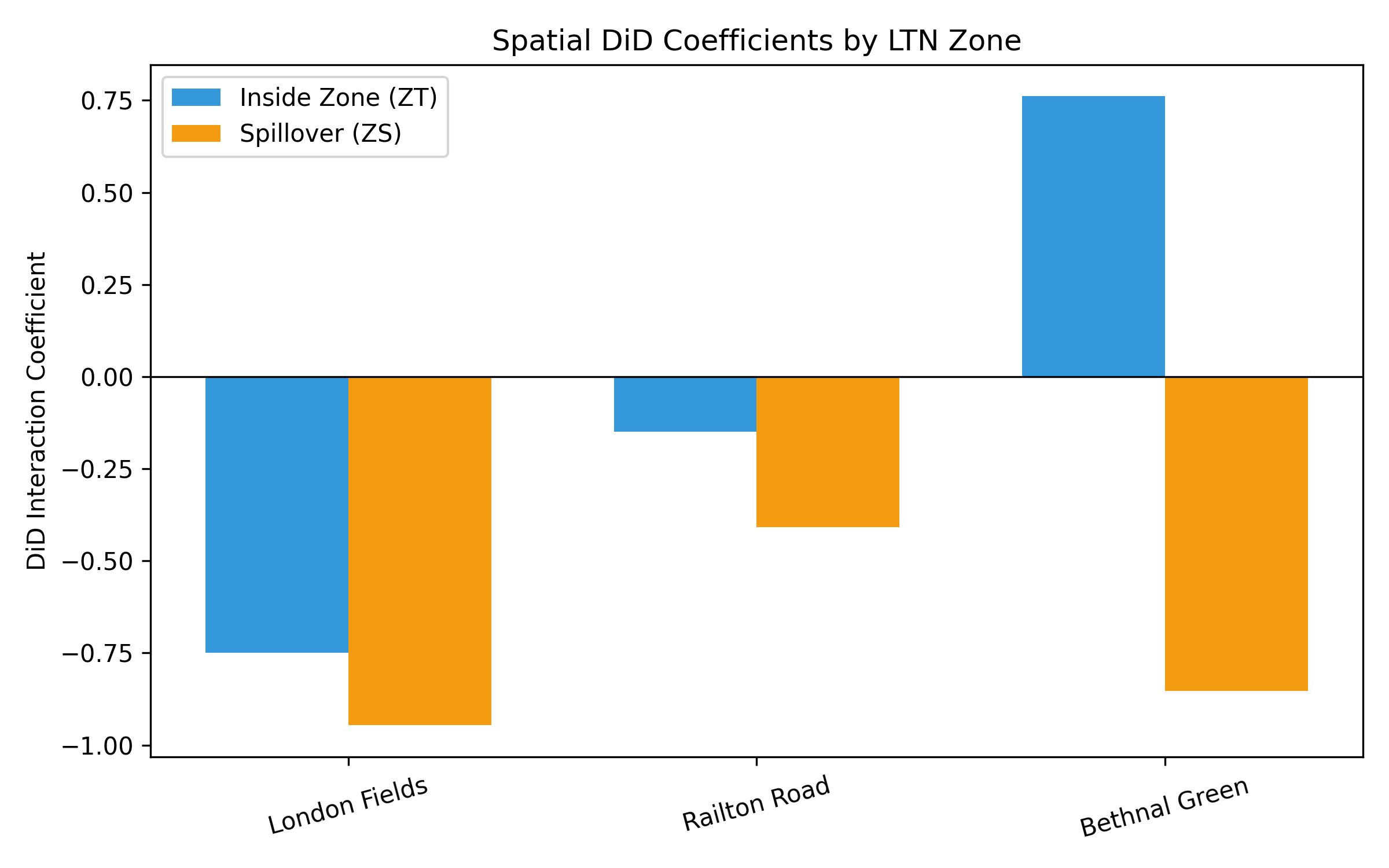}
    \caption{Spatial DiD interaction coefficients for the three LTN zones with sufficient pre-period data for EB/DiD estimation. Separate $Z_T$ and $Z_S$ estimates show mixed inside-zone effects and no consistent spillover worsening.}
    \label{fig:ltn_did}
\end{figure}

\subsubsection{Cost--Benefit Analysis: Break-Even Threshold}

\begin{figure}[htbp]
    \centering
    \includegraphics[width=0.85\textwidth]{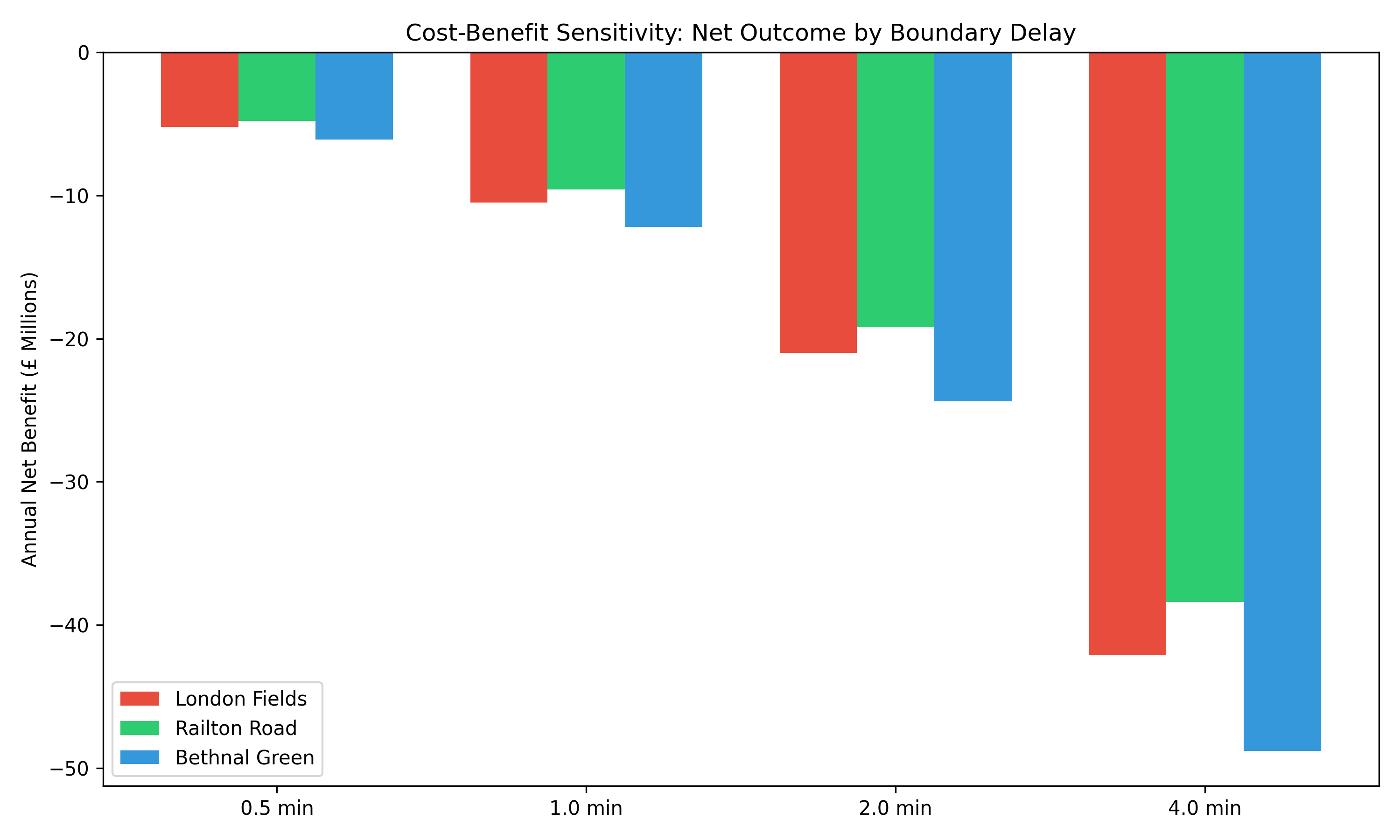}
    \caption{Break-even threshold sensitivity chart. Under the assumed delay scenarios, even modest boundary-road delays produce negative net benefits. The critical question is whether observed delays exceed the break-even threshold of approximately 10~seconds per vehicle.}
    \label{fig:ltn_cba}
\end{figure}

Because no clear casualty reduction was detected
across most zones (e.g., London Fields: $\Delta$Serious~$= -1$,
$\Delta$Slight~$= -4$), the break-even point for journey-time costs is
low. Figure~\ref{fig:ltn_cba} and Table~\ref{tab:cba_grid}
(Appendix~\ref{app:cba_grid}) present the sensitivity analysis. Under
the assumed delay of 2~minutes per vehicle, even modest boundary-road
traffic volumes generate annualised costs that exceed the monetised safety
gains. At the base-case AADF of 15,000, we calculate a break-even
threshold of approximately 10~seconds of average delay per vehicle on
boundary roads.\footnote{%
Break-even arithmetic: London Fields EB-corrected
$\Delta$Serious~$= -1$, $\Delta$Slight~$= -4$. Monetised safety benefit
$= 1 \times \text{\pounds}220{,}434 + 4 \times \text{\pounds}20{,}000
= \text{\pounds}300{,}434$/year. Annual boundary vehicles $= 15{,}000
\times 365 = 5{,}475{,}000$. At VTT $= \text{\pounds}20$/hour
($= \text{\pounds}0.00556$/s), break-even delay
$= \text{\pounds}300{,}434 / (5{,}475{,}000 \times 0.00556)
\approx 9.9$~seconds per vehicle.}
If boundary delays exceed this threshold, the monetised journey-time
costs would negate the safety gains. Conversely, if delays remain below
it, the economic case would be more favourable. Resolving this question
requires empirical journey-time data, which are not available in the
current study.

\subsection{\texorpdfstring{20\,mph}{20 mph} Speed Limit Results}

Before examining conditional severity shares, we report the primary
count-based outcomes. During 2020--2024, London recorded 53,434 collisions on
20\,mph roads (8,515 KSI) and 49,728 on 30\,mph roads (7,630 KSI).
Annualised per-borough KSI counts are higher in 20\,mph-adopted boroughs (mean
243/year vs.\ 188/year in controls), but 20\,mph-adopted boroughs also have higher
total collision volumes (mean 1,534/year vs.\ 1,190/year), reflecting their
denser, more complex road environments rather than a speed-limit effect.

\subsubsection{Borough-Level Comparison}

20\,mph-adopted boroughs show a higher median KSI severity share (17.8\%) than
30\,mph-default boroughs (14.9\%), a difference that is statistically
significant ($p < 0.001$, Mann--Whitney~$U$). However, this comparison is
heavily confounded by inner-London characteristics: 20\,mph-adopted boroughs
are all dense, inner-city areas with more pedestrians, cyclists, narrower
streets, and more junctions per km than the suburban outer-London controls.
The association should not be interpreted as evidence that
20\,mph limits \emph{increase} severity.

\paragraph{The confound reinforces the policy-fit critique.}
Critics may argue that this confound invalidates the comparison entirely.
We argue instead that it reinforces the central policy critique: even if the
higher KSI share on 20\,mph A-roads is driven entirely by the fact that
inner-city corridors have vastly higher vulnerable road user exposure,
\emph{this underscores exactly why sign-only defaults are structurally
insufficient on these roads}. If these are precisely the environments with
the greatest concentration of serious conflict, then a cheap statutory
default is the weakest possible policy response. These corridors demand
physical engineering, protected crossings, and active enforcement, rather than a
blanket speed limit that drivers routinely ignore.

\paragraph{Exposure-rate data gap.} Absolute KSI rates per million vehicle-km
by speed-limit group and road class would supplement the compositional
analysis with exposure-based evidence. However, segment-level AADF data
linked to speed-limit classification are not currently available. This is a
priority for the next iteration of this working paper.

\paragraph{Methodological note.} A stronger approach to this confound would
require Propensity Score Matching (PSM) at the road-segment level as a descriptive robustness check,
matching inner-London 20\,mph segments with comparable inner-London 30\,mph
segments on road geometry, traffic volume, VRU exposure, and junction density.
More robust next-step designs would include adoption-period panels, segment-level fixed-effects DiD, synthetic control, and exposure-based count models. These remain priorities for future work.

\begin{figure}[htbp]
    \centering
    \includegraphics[width=0.85\textwidth]{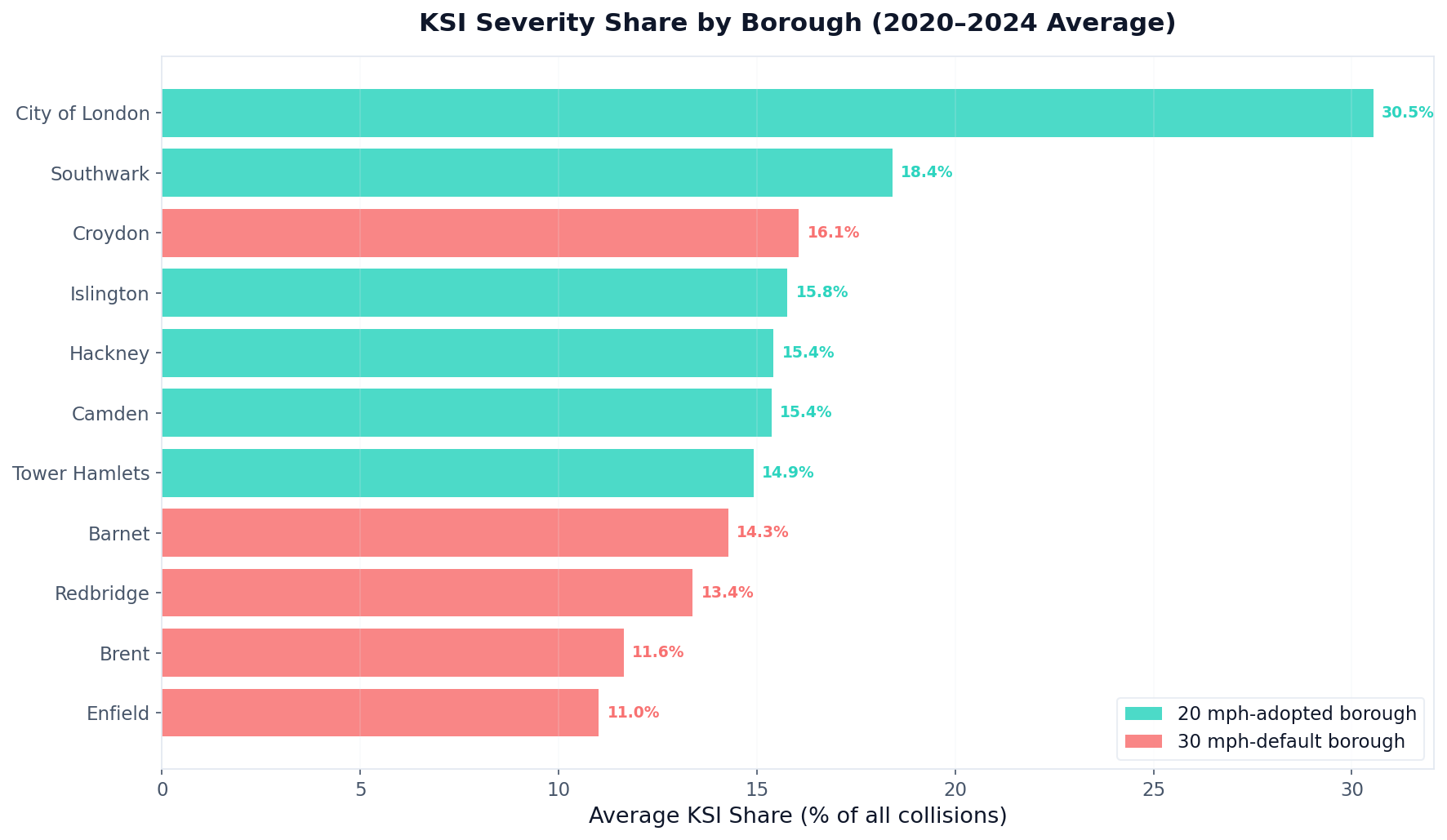}
    \caption{KSI share by borough. 20\,mph-adopted boroughs are all in inner London with fundamentally different road layouts than 30\,mph-default boroughs.}
    \label{fig:mph_borough}
\end{figure}

\subsubsection{Logistic Regression (Severity Conditional on Collision)}

We report results under two specifications. \textbf{Specification~A}
(environment-only: 9 pre-crash covariates plus year) yields
OR~=~1.051 (95\% CI: 1.015--1.088, $p = 0.005$). \textbf{Specification~B}
(14 covariates including post-crash controls \texttt{number\_of\_vehicles}
and \texttt{number\_of\_casualties}) yields OR~=~1.055 (95\% CI:
1.019--1.092, $p = 0.003$). The difference between specifications is
negligible ($\Delta$OR $< 0.4\%$), suggesting the post-crash controls do
not substantially drive the result. However, both estimates are subject to
at least three identification threats:

\begin{enumerate}
    \item \textbf{Over-control / collider bias:} The inclusion of
          \texttt{number\_\allowbreak of\_\allowbreak vehicles} and \texttt{number\_\allowbreak of\_\allowbreak casualties}
          in Spec~B blocks potential mediation pathways and may open non-causal
          back-door paths (see \S\ref{sec:logit}).
    \item \textbf{Selection on the outcome:} Both specifications condition on a
          collision existing; they cannot inform us about collision \emph{risk}.
    \item \textbf{Residual confounding:} 20\,mph roads are disproportionately
          in dense inner-London areas with higher pedestrian and cyclist
          exposure, features only partially captured by the covariates.
          A more robust check would involve Propensity Score Matching (PSM)
          at the road-segment level. However, stronger subsequent designs, such as adoption-period panels, segment-level fixed-effects DiD, synthetic controls, and exposure-based count models, are required for more definitive insights.
\end{enumerate}

These figures represent the \emph{conditional association between the
speed-limit regime and severity composition}, not an estimate of the causal
effect of adopting a 20\,mph limit. Moreover, with $N = 103{,}162$, an
odds ratio of approximately 1.05 is statistically significant but
practically small. An effect of this magnitude is difficult to distinguish
from residual confounding in an observational design and should be
interpreted as a small associational signal rather than strong evidence of
a policy effect.

\begin{table}[htbp]
    \centering
    \caption{Logistic regression results: KSI outcome on 20\,mph vs.\ 30\,mph roads.}
    \label{tab:logit}
    \begin{tabular}{llrrrr}
        \toprule
        \textbf{Specification} & & \textbf{OR} & \textbf{95\% CI} & \textbf{$p$} & \textbf{$N$} \\
        \midrule
        A (environment-only) & Speed 20\,mph & 1.051 & [1.015, 1.088] & 0.005 & 103,162 \\
        B (predictive)       & Speed 20\,mph & 1.055 & [1.019, 1.092] & 0.003 & 103,162 \\
        \bottomrule
    \end{tabular}
\end{table}

\begin{figure}[htbp]
    \centering
    \includegraphics[width=0.85\textwidth]{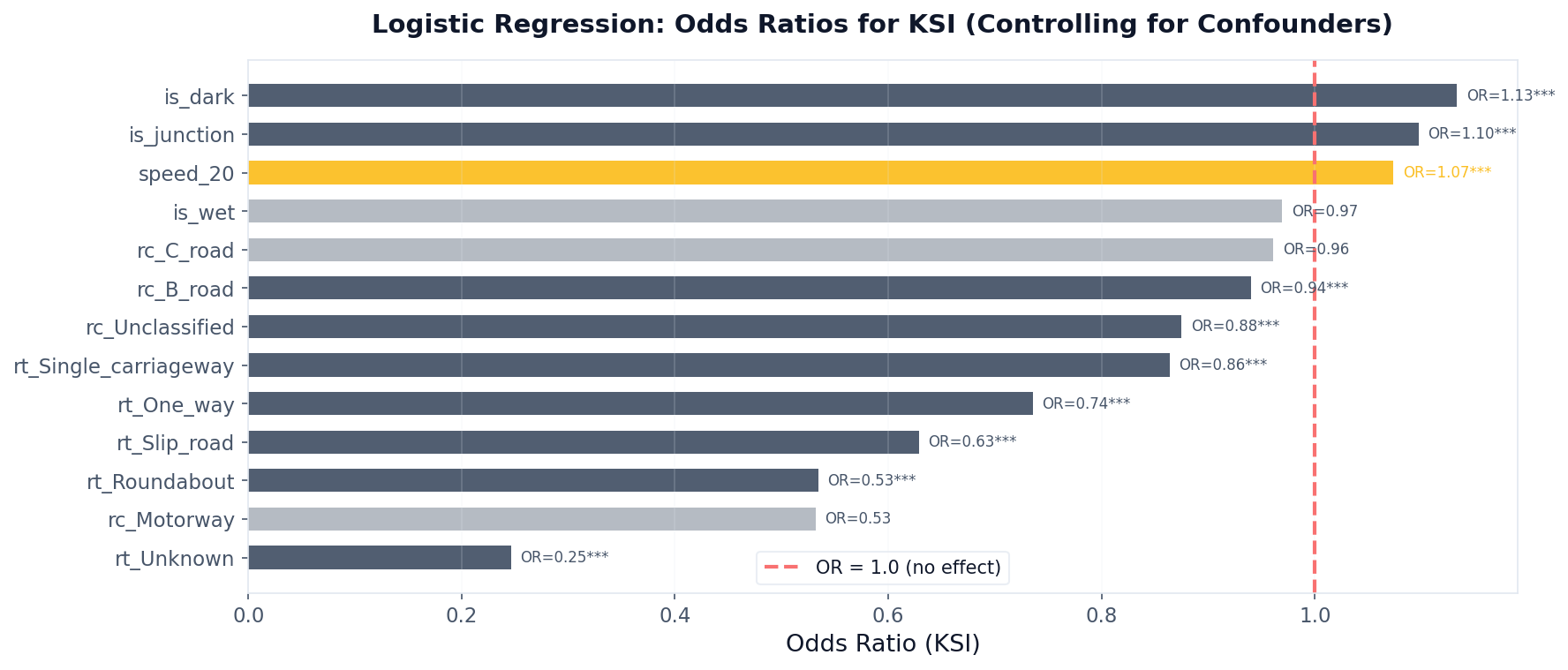}
    \caption{Odds ratios from the logistic regression model. The speed limit variable (speed\_20) is highlighted.}
    \label{fig:mph_logistic}
\end{figure}

\subsubsection{Stratified Analysis by Road Class}

When comparing 20mph and 30mph roads within the same road class, the picture becomes nuanced (Table~\ref{tab:strat}).

\begin{table}[htbp]
    \centering
    \caption{KSI share difference (20mph $-$ 30mph) within each road class stratum.}
    \label{tab:strat}
    \resizebox{\linewidth}{!}{%
    \begin{tabular}{lrrrl}
        \toprule
        \textbf{Road Class} & \textbf{KSI 20mph} & \textbf{KSI 30mph} & \textbf{$\Delta$ (pp)} & \textbf{Association} \\
        \midrule
        A road (arterial) & 17.4\% & 15.6\% & $+1.78$ & Higher KSI share (assoc.) \\
        B road (secondary) & 15.6\% & 15.2\% & $+0.44$ & No clear difference (assoc.) \\
        C road (residential) & 15.2\% & 15.7\% & $-0.44$ & Lower point est.\ (not robust after FDR) \\
        Unclassified (local) & 13.5\% & 14.3\% & $-0.86$ & Lower point est.\ (not robust after FDR) \\
        \bottomrule
    \end{tabular}%
    }
\end{table}

\subsubsection{Spatial Boundary Comparison}

A spatial boundary comparison was attempted using two borough-boundary pairs
(Camden--Barnet and Camden--Brent), but the design lacks statistical power
(2 pairs, $\chi^2 = 1.25$, $p = 0.264$). Full results are reported in
Appendix~\ref{app:boundary}.

\subsubsection{Heterogeneous Associations}

\begin{figure}[htbp]
    \centering
    \includegraphics[width=0.95\textwidth]{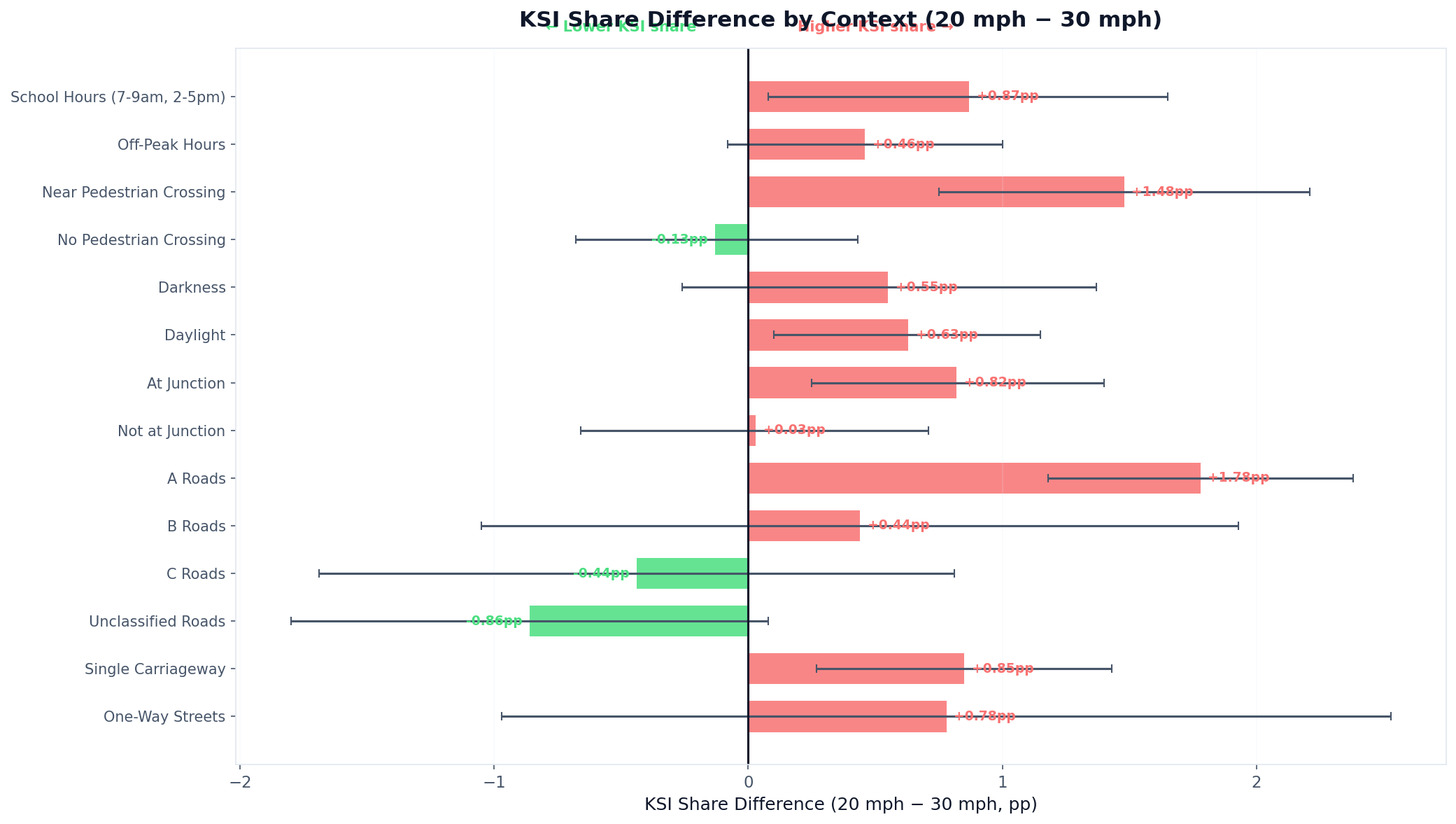}
    \caption{KSI share difference (20\,mph $-$ 30\,mph) by context, with 95\% confidence intervals. Bars to the left indicate a lower KSI share on 20\,mph roads; bars to the right indicate a higher share.}
    \label{fig:het}
\end{figure}

We evaluate heterogeneity across 14 pre-specified contexts by comparing KSI
share between 20\,mph and 30\,mph roads within each context.
Figure~\ref{fig:het} plots the estimated difference in KSI share
(20\,mph~$-$~30\,mph) with uncertainty intervals, and Table~\ref{tab:het}
reports false discovery rate (FDR) adjusted $p$-values to account for
multiple comparisons.

Point estimates vary across contexts, with some residential settings showing
small negative differences (e.g., C-roads, unclassified roads). However, these
negative differences are not statistically robust once we control the FDR
across the full set of subgroup tests. In contrast, four contexts show robustly
higher KSI shares on 20\,mph roads after FDR adjustment: A-roads
($+1.78$\,pp), near pedestrian crossings ($+1.48$\,pp), at junctions
($+0.82$\,pp), and single carriageways ($+0.85$\,pp). This pattern suggests
that posted 20\,mph limits are not associated with lower KSI severity
composition in the higher-risk environments where many serious outcomes occur
(arterials and junction-dense locations).

These subgroup comparisons remain associational and are conditional on an
injury collision being recorded in STATS19; they should therefore be
interpreted as descriptive evidence of where 20\,mph and 30\,mph collision
profiles differ, rather than definitive causal effects of the policy.

\begin{table}[htbp]
    \centering
    \caption{Heterogeneous associations by context: KSI share for 20\,mph vs.\ 30\,mph roads, difference
    (20\,mph$-$30\,mph, percentage points), and FDR-adjusted $p$-values.}
    \label{tab:het}
    \small
    \begin{tabular}{lrrrrrrl}
        \toprule
        \textbf{Context} & \textbf{$N_{20}$} & \textbf{$N_{30}$}
            & \textbf{KSI$_{20}$} & \textbf{KSI$_{30}$}
            & \textbf{$\Delta$pp} & \textbf{$p_{\text{FDR}}$}
            & \textbf{Sig.} \\
        \midrule
        A Roads           & 27,886 & 30,802 & 17.36\% & 15.58\% & $+1.78$ & $<0.001$ & Yes \\
        Near Ped.\ Crossing & 22,115 & 19,001 & 17.83\% & 16.35\% & $+1.48$ & $<0.001$ & Yes \\
        Single Carriageway  & 33,826 & 31,989 & 17.58\% & 16.72\% & $+0.85$ & $0.018$ & Yes \\
        At Junction         & 33,981 & 30,539 & 17.22\% & 16.40\% & $+0.82$ & $0.018$ & Yes \\
        \midrule
        School Hours        & 16,342 & 15,564 & 15.56\% & 14.69\% & $+0.87$ & $0.072$ & No \\
        One-Way Streets     & 5,075  & 2,302  & 15.29\% & 14.51\% & $+0.78$ & $0.538$ & No \\
        Daylight            & 36,944 & 34,055 & 15.39\% & 14.77\% & $+0.63$ & $0.056$ & No \\
        Darkness            & 16,490 & 15,673 & 17.15\% & 16.60\% & $+0.55$ & $0.287$ & No \\
        Off-Peak Hours      & 37,092 & 34,164 & 16.10\% & 15.64\% & $+0.46$ & $0.162$ & No \\
        B Roads             & 5,543  & 3,822  & 15.64\% & 15.20\% & $+0.44$ & $0.657$ & No \\
        Not at Junction     & 19,453 & 19,189 & 13.69\% & 13.67\% & $+0.03$ & $0.942$ & No \\
        No Ped.\ Crossing   & 31,319 & 30,727 & 14.60\% & 14.72\% & $-0.13$ & $0.710$ & No \\
        C Roads             & 6,311  & 6,574  & 15.24\% & 15.68\% & $-0.44$ & $0.624$ & No \\
        Unclassified        & 13,690 & 8,476  & 13.48\% & 14.33\% & $-0.86$ & $0.144$ & No \\
        \bottomrule
    \end{tabular}
\end{table}

\paragraph{Counts companion.}
Context decomposition in this section is only feasible for the KSI severity
share; count-based context models would require segment-time exposure data
that are not available in the current STATS19 extract. For the aggregate
count-based comparison, see Table~\ref{tab:count_outcomes} and \S\ref{sec:results}.

\subsubsection{Speed--Safety Curve}

\begin{figure}[htbp]
    \centering
    \includegraphics[width=0.85\textwidth]{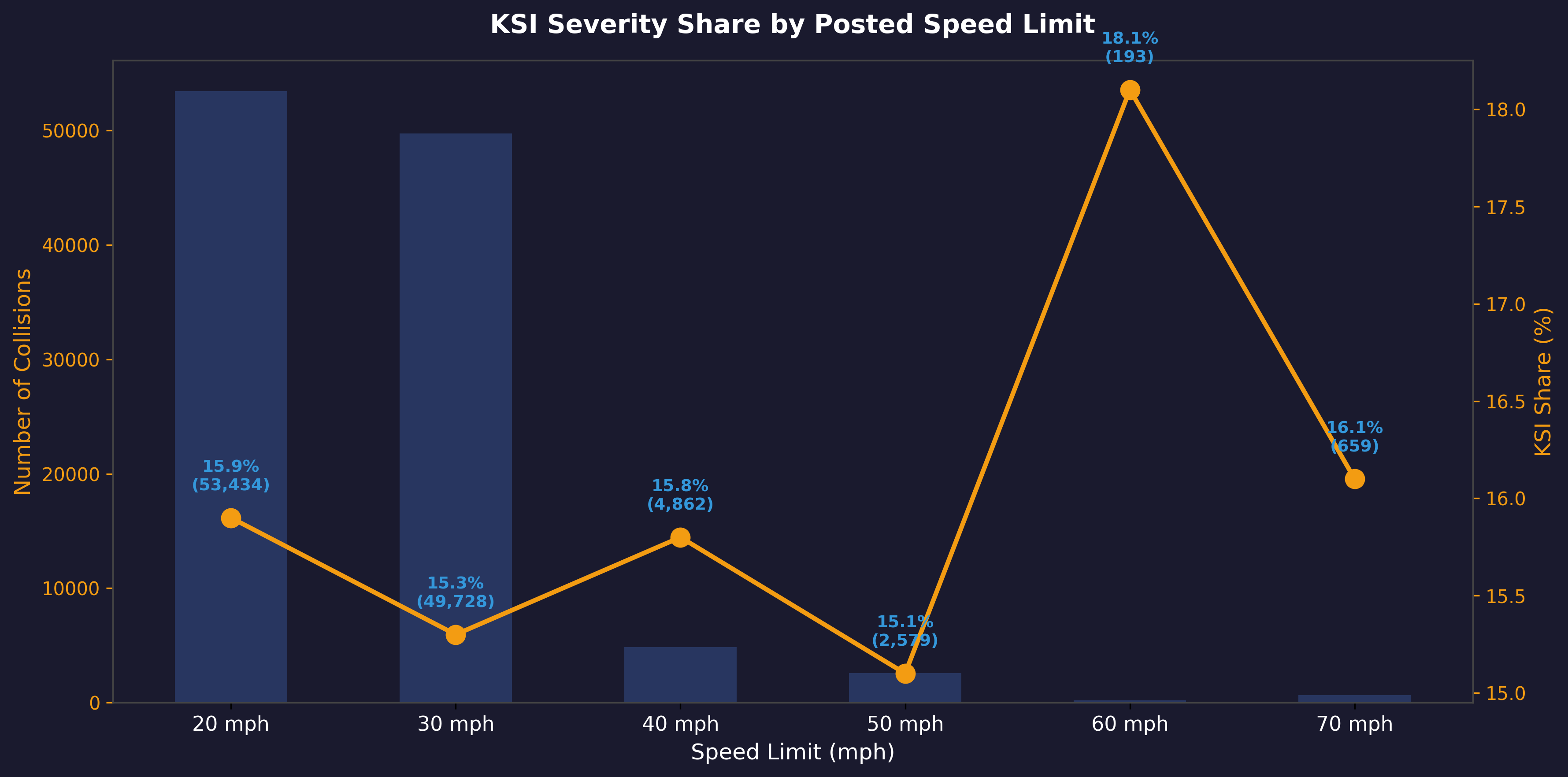}
    \caption{Posted-limit vs.\ KSI-share descriptive profile. The KSI share difference between 20\,mph and 30\,mph is only 0.6 percentage points.}
    \label{fig:speed_curve}
\end{figure}

Table~\ref{tab:curve} and Figure~\ref{fig:speed_curve} present the
descriptive association in aggregates between posted speed limit and KSI share. The posted-limit
vs.\ KSI-share association is weak in these observational aggregates,
consistent with strong confounding by road type, limited compliance
differences in congested inner London, and the compositional nature of
the share measure.

\begin{table}[htbp]
    \centering
    \caption{KSI share by posted speed limit across all London collisions.}
    \label{tab:curve}
    \begin{tabular}{rrr}
        \toprule
        \textbf{Speed Limit} & \textbf{Collisions} & \textbf{KSI Share} \\
        \midrule
        20 mph & 53,434 & 15.9\% \\
        30 mph & 49,728 & 15.3\% \\
        40 mph & 4,862 & 15.8\% \\
        50 mph & 2,579 & 15.1\% \\
        60 mph & 193 & 18.1\% \\
        70 mph & 659 & 16.1\% \\
        \bottomrule
    \end{tabular}
\end{table}

\section{Discussion}\label{sec:discussion}

\subsection{LTNs: Limited Evidence of Casualty Reduction, 2020--2024}

Our three-step pipeline reveals that na\"ive pre/post estimates dramatically
overstate apparent LTN effects due to RTM. After EB correction, only one
zone (London Fields) retains a statistically significant reduction in
collision counts in the Spatial DiD.

\paragraph{No consistent crash displacement: why this matters.}
Crucially, we do not observe a consistent pattern of positive spillover
estimates that would be expected under systematic crash displacement. We find
no systematic evidence that LTNs in our sample displaced crashes onto boundary
roads over 2020--2024. However, the data \emph{equally} do not support the
sweeping inside-zone safety benefits that proponents frequently assert: after
RTM correction, the claimed safety gains largely evaporate.

This finding does not imply that LTNs have \emph{no} benefits: they may
reduce air pollution, noise, and improve residential amenity inside the zone.
However, in our sample over 2020--2024, the road-safety claims frequently
advanced in their favour are not supported uniformly when RTM and
network-wide effects are properly accounted for.

\paragraph{Break-even threshold analysis.}
Because the EB-corrected safety benefits represent small estimated casualty deltas across
most zones (e.g., London Fields saw an estimated reduction of just 1 serious
and 4 slight casualties), the break-even point for journey-time delay is
low. We calculate that an average journey-time
increase of 10~seconds per vehicle on boundary roads at a base-case AADF
of 15,000 would negate the monetised economic value of these safety gains.
We note that journey-time delays must be measured empirically to
resolve this economic question in either direction. In the absence
of such evidence, the cost--benefit case for LTNs as a safety intervention
remains unproven.

\subsection{\texorpdfstring{20\,mph}{20 mph} Regimes: Sign-Only Defaults and Road-Function Mismatch}\label{sec:disc-20mph}

The 20\,mph analysis reveals associational patterns that vary substantially
across road environments, and the pattern of variation is precisely what
a road-function mismatch critique of blanket sign-only/default regimes would predict.
When decomposed by context:

\begin{itemize}
    \item \textbf{Negative point estimates (not FDR-robust):}
          C-roads ($-0.44$\,pp, $p_{\text{FDR}} = 0.624$), unclassified roads
          ($-0.86$\,pp, $p_{\text{FDR}} = 0.144$), and areas away from
          pedestrian crossings ($-0.13$\,pp, $p_{\text{FDR}} = 0.710$).
          These correspond to quieter residential settings where lower speeds
          plausibly reduce injury severity and where sign-only treatment may
          be closer to self-enforcing. However, the evidence is suggestive
          rather than statistically confirmed after multiple-testing correction.
    \item \textbf{FDR-robust increases in KSI share on 20\,mph roads:}
          A-roads ($+1.78$\,pp, $p_{\text{FDR}} < 0.001$), at junctions
          ($+0.82$\,pp, $p_{\text{FDR}} = 0.018$), near pedestrian crossings
          ($+1.48$\,pp, $p_{\text{FDR}} < 0.001$), and single carriageways
          ($+0.85$\,pp, $p_{\text{FDR}} = 0.018$).
    \item \textbf{Non-robust positive point estimates:}
          B-roads ($+0.44$\,pp), school hours ($+0.87$\,pp), and darkness
          ($+0.55$\,pp) show positive point estimates that do not survive
          FDR correction.
\end{itemize}

\paragraph{Road-function mismatch and strategic-road misapplication.}
The four FDR-robust positive contexts (A-roads, junctions, pedestrian
crossings, and single carriageways) share a common feature: they are
precisely the corridors where a sign-only 20\,mph limit is structurally
mismatched to road function. A-roads are designed for higher
traffic speeds and flows; junctions concentrate conflict points and complex
manoeuvres; pedestrian crossings attract vulnerable road users to locations
with mixed traffic. On these large roads whose primary role is movement rather than place, often with limited crossing activity, weak frontage, and geometry that cues higher operating speeds, signage alone is unlikely to self-enforce. DfT's own evaluation found that
sign-only schemes produced median speed reductions of less than
1\,mph \citep{DfT2018Headline}, and \citet{Quddus2024} confirm that
sign-only schemes deliver materially weaker safety outcomes than schemes
with physical calming. The most plausible reading of our findings is not that lower urban speeds fail in principle, but that blanket, largely sign-only/default 20\,mph is a weak and poorly targeted treatment on heterogeneous inner-London roads, especially on strategic corridors and conflict-heavy arterial environments.

\paragraph{The compositional trap and Vision Zero objectives.}
Critics will note that the KSI severity share is compositional: a higher share
on 20\,mph A-roads could simply mean that these limits successfully reduced
minor ``slight'' collisions while leaving fatal and serious collisions unchanged.
We acknowledge this possibility explicitly (see \S\ref{sec:outcomes}). But
even under this most charitable interpretation, the policy implications for
blanket 20\,mph limits are challenging. If the compositional reading is correct,
it would mean that sign-only defaults on strategic arterials impose systemic
journey-time costs and recurring operating burdens (affecting bus operations, freight
logistics, and commuters) merely to prevent low-severity collisions (fender
benders and minor scrapes), while failing to reduce the fatal and serious
injuries that 20\,mph limits were explicitly implemented to stop. A policy
that slows strategic corridors at substantial economic cost but is not associated
with measurable reductions in serious harm would be failing to address Vision Zero
objectives on its own terms.

\noindent Exposure-rate analysis (KSI per million vehicle-km by speed-limit
group and road class) is not possible with the current dataset because
segment-level AADF data are not linked to speed-limit classification.
This is a priority for the next iteration.

\paragraph{Compliance and the recurring burden of the wrong-road problem.}
When a statutory limit actively mismatches the functional character of a road, compliance collapses. DfT Circular 01/2013 stipulates that `successful 20\,mph schemes are generally self-enforcing' \citep{DfT2013}. On arterial corridors, self-enforcement is a fiction; achieving compliance requires intensive, perpetual police enforcement at severe recurring cost. The Welsh national monitoring report documented widespread journey-time increases on through-routes following the September 2023 default \citep{TfW2025}. 

These recurring operating costs, including bus schedule disruption, freight delays on strategic corridors, and systemic journey-time drag on through-traffic, are not trivial. They compound massively across millions of trips. The strongest economic critique of blanket default 20\,mph is not that it triggers an apocalyptic collapse of local economies, but that it imposes persistent, structural drag on strategic corridors where the safety payoff of sign-only treatment is weak or non-existent.

\paragraph{Public sentiment and the democratic legitimacy of blanket defaults.}
The Welsh 20\,mph default provides an instructive case study of how blanket speed-limit policy interacts with democratic consent. The Welsh Government's formal consultation on the proposed default received approximately 6,000 analysed responses: 53\% opposed the measure and 47\% supported it \citep{WelshConsultation2022}. Because public consultations are self-selecting, this split alone is not decisive. However, a representative omnibus survey commissioned alongside the consultation found stronger in-principle support for 20\,mph on residential streets. This suggests the public was not opposed to lower speeds per se, but rather to the indiscriminate application of a default across all restricted roads, including movement-function corridors that respondents judged unsuitable.

The distinction became critical after implementation. Post-rollout polling by YouGov recorded overwhelming opposition to the national default once drivers experienced it in practice, along with widespread self-reported non-compliance \citep{YouGov2024}. This trajectory, with initial in-principle support collapsing into active hostility after deployment, is not evidence that 20\,mph limits are unsafe. It is, however, powerful evidence of a \emph{legitimacy, targeting, and communication failure}. When a statutory limit is perceived as misapplied to roads where the restriction contradicts the road's physical character, public consent erodes rapidly. Crucially, this erosion is not confined to the mismatched roads: it poisons compliance and political support for speed management across the entire network, including on residential streets where a lower limit genuinely belongs.

The Welsh Government's subsequent exception review returned hundreds of arterial and through-route sections to 30\,mph \citep{WelshExceptions2024}, an implicit admission that the blanket default was poorly targeted. DfT's own sign-only evaluation in England documented similar dynamics at a smaller scale: driver frustration with limits perceived as unreasonable, and bus operators voicing concern that blanket limits degraded timetable reliability without demonstrable safety gains \citep{DfT2018Technical}. When sustained voter dissatisfaction, self-reported non-compliance, and political reversal all converge on the same class of roads (strategic, movement-function corridors), which provides strong circumstantial evidence of poor policy fit, not mere public obstinacy.

\paragraph{Mechanistic hypotheses.}
Several mechanisms \emph{could} explain the higher severity shares on busier
20\,mph roads. We frame these as hypotheses for future investigation, not as
confirmed findings:

\begin{enumerate}
    \item \textbf{Speed variance hypothesis:} On arterial roads designed for
          higher speeds, a 20\,mph limit may produce high speed variance (where some
          drivers comply while others do not), which is associated with elevated
          crash risk in the traffic safety literature.
    \item \textbf{Maneuverability hypothesis:} At merging points, 20\,mph may
          reduce the acceleration headroom needed for smooth zipper merging,
          potentially increasing conflict frequency.
    \item \textbf{Prolonged-exposure hypothesis:} A narrower speed differential
          between cars and bicycles might extend overtaking durations, increasing
          the time cyclists spend in vehicle blind spots.
\end{enumerate}

We cannot test these mechanisms without observed speed distributions and compliance data.

\paragraph{Descriptive evidence from STATS19.}
Consistent with (but not confirming) these hypotheses, we observe that
collisions on 20\,mph arterial roads are more concentrated at junctions and
merging points (76.6\% vs.\ 70.6\% on 30\,mph arterials) and that VRU
collisions have a slightly higher KSI severity share on 20\,mph arterials
(24.8\% vs.\ 22.8\%). These are \emph{associational patterns} that could
reflect the mechanisms above, but also selection effects: 20\,mph arterials
may be located in locations with inherently higher conflict rates. Testing
the causal pathway would require direct measures of actual driving speeds,
speed variance, and overtaking behaviour, which are not available in STATS19.

\begin{figure}[htbp]
    \centering
    \includegraphics[width=0.85\textwidth]{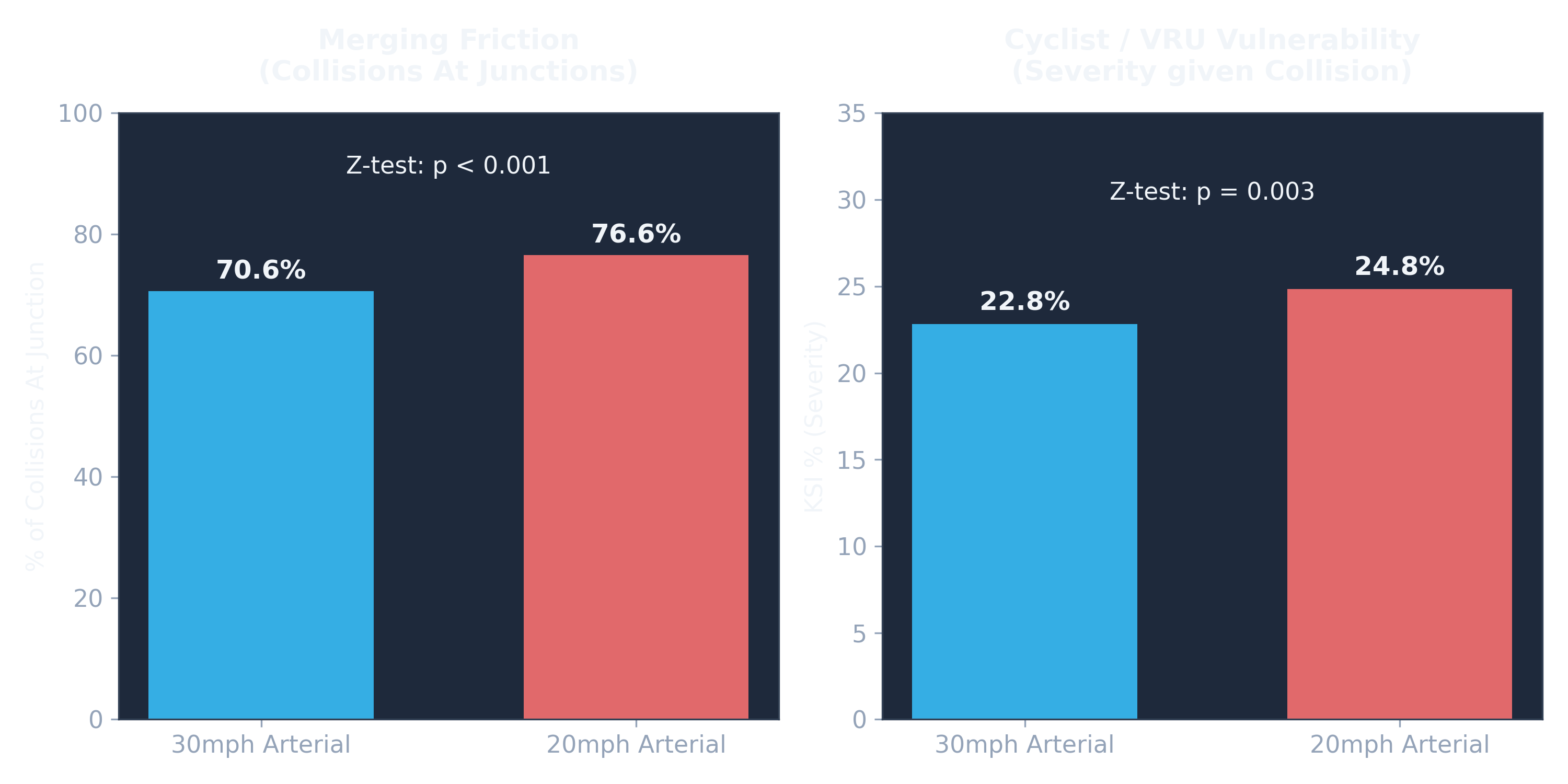}
    \caption{Associational patterns: collision concentration at junctions and VRU severity on 20\,mph vs.\ 30\,mph arterial roads. These are descriptive patterns consistent with the hypotheses in \S\ref{sec:disc-20mph}, not causal evidence.}
    \label{fig:mph_maneuver}
\end{figure}

The speed--severity-share curve's flatness between 20--50\,mph in London is
noteworthy. One explanation is that actual driving speeds may not differ
substantially between 20\,mph and 30\,mph zones, particularly in congested
inner London where speeds rarely reach 30\,mph regardless of the posted limit.
This is consistent with DfT's finding that sign-only schemes achieve median
speed reductions of less than 1\,mph \citep{DfT2018Headline}: if actual
speeds are barely affected, there is little mechanism through which a
changed posted limit can alter collision outcomes.

\subsection{Limitations}\label{sec:limitations}

Several important caveats apply to this study. We organise them by severity:

\paragraph{Data limitations.}
\begin{itemize}
    \item STATS19 records only police-reported injury collisions. Minor
          incidents and damage-only collisions are not captured, which may bias
          the KSI severity share upward (if slight collisions are
          disproportionately underreported) and understate total collision
          counts.
    \item The pre-intervention period for LTNs introduced in mid-2020 is
          limited to January--June 2020, a period severely distorted by
          COVID-19 lockdowns. Traffic volumes during lockdown were 30--70\%
          below normal levels, rendering this baseline unrepresentative of
          normal operating conditions. This distortion does not merely affect
          our analysis: it makes na\"ive pre/post comparisons throughout the
          broader LTN evaluation literature, including the widely cited
          \citet{Laverty2021} and \citet{Aldred2021} estimates, highly
          suspect whenever they rely on 2020 as a baseline year. The
          necessary next step for this working paper is to integrate STATS19
          data from 2017--2019 to establish a clean, non-pandemic
          pre-intervention baseline, providing the statistical power needed
          to run an Event-Study specification and definitively test
          parallel trends.
    \item LTN zone boundaries are approximations based on published
          coordinates, not official council GIS shapefiles. Misclassification
          of collisions near boundaries could attenuate estimates.
\end{itemize}

\paragraph{Identification limitations.}
\begin{itemize}
    \item \textbf{KSI severity share $\neq$ safety risk:} Our primary
          reported outcome for the 20\,mph analysis (the KSI severity
          share) conditions on a collision existing and being reported. It
          captures severity composition, not the probability of being killed
          or seriously injured per trip or per km. Conclusions about safety
          require count-based outcomes and exposure denominators (see
          \S\ref{sec:outcomes}).
    \item \textbf{No within-window policy variation for 20\,mph:} All
          borough-wide 20\,mph adoptions occurred before 2020. Our comparisons
          are cross-sectional associations at best (\S\ref{sec:20mph-id}).
    \item \textbf{Logistic regression conditions on collision:} Even the
          environment-only specification cannot identify the causal effect on
          collision risk. It estimates the association with severity composition
          conditional on a crash having already occurred (\S\ref{sec:logit}).
    \item \textbf{Over-control / collider bias:} Including post-crash
          covariates (vehicles, casualties) in the predictive specification
          blocks mediation pathways and may open non-causal paths.
    \item \textbf{Spatial boundary comparison is not an RDD:} With only 2
          boundary pairs and no running variable, the design lacks the
          statistical power and local-identification properties of a genuine
          regression discontinuity (\S\ref{sec:boundary}).
    \item \textbf{LTN parallel trends not verified:} The DiD assumes parallel
          trends, which we have not yet tested via an event-study specification.
    \item \textbf{EB prior is unconditional:} The Empirical Bayes correction
          does not condition on site-specific characteristics beyond the zone
          classification (\S\ref{sec:eb}).
\end{itemize}

\paragraph{Inference limitations.}
\begin{itemize}
    \item \textbf{Multiple testing:} The heterogeneous associations analysis tests
          14 overlapping contexts. After Benjamini--Hochberg correction, some
          nominally significant differences may lose significance.
    \item \textbf{Serial correlation and time aggregation:} LTN data are
          aggregated into two periods (pre/post) rather than a panel of
          monthly or quarterly observations. This coarse temporal resolution
          limits the precision of DiD estimates and precludes event-study
          validation of parallel trends. Because the pre-period is short and
          COVID-distorted for mid-2020 LTNs, an event study is unlikely to
          provide reliable pre-trend evidence in this window; we treat it as
          a planned extension with earlier-year data.
    \item \textbf{CBA is assumption-driven:} The cost--benefit analysis
          depends heavily on the assumed per-vehicle delay, which is not
          directly measured. The net-benefit conclusion could reverse under
          different delay assumptions (\S\ref{sec:cba}).
\end{itemize}

\subsection{Policy Context: Environmental Externalities: Emissions, Congestion, and the Paradox of Arterial Speed Reduction}\label{sec:emissions}
The emissions impact of 20\,mph limits is contested and context-dependent. The following presents broader policy context beyond the direct outputs of our 2020--2024 collision design. The balance of evidence suggests that while 20\,mph can provide environmental benefits in specific settings, blanket application to arterial through-routes is likely to be counter-productive or neutral at best.

There is a robust engineering argument that 20\,mph limits reduce certain emissions by capping the energy required for acceleration. Because kinetic energy scales with the square of velocity ($E_k = \frac{1}{2}mv^2$), accelerating a vehicle from rest to 30\,mph requires 2.25 times the energy of accelerating to 20\,mph. In heavily trafficked urban centres characterised by frequent stop-start cycles, a lower peak speed reduces energy waste per cycle. Modelling by Future Transport (2021) suggests this dynamic could yield CO$_2$ reductions of $\sim$26\% and NO$_x$ reductions of $\sim$28\% in realistic urban drive cycles. Furthermore, a City of London evaluation by Imperial College \citep{Williams2013} found that smoother driving profiles at 20\,mph reduced tyre and brake particulate wear, concluding that it would be incorrect to assume 20\,mph is systematically detrimental to air quality.

However, the empirical literature and mechanical reality of internal combustion engine (ICE) vehicles present a more mixed picture. A recent systematic review found no consensus on the direction or magnitude of emission changes from low-speed limit interventions, noting high variability by location and condition \citep{MDPISustainability2024}. At steady cruise speeds, ICE vehicles are generally less fuel-efficient at 20\,mph than 30\,mph because they are forced to operate in lower gears at suboptimal engine RPM. Driving experts note this can increase per-distance fuel consumption by up to 50\% on free-flowing roads. Indeed, the City of London study itself noted that for larger vehicles ($>2.0$\,L engines), emissions per km could actually increase at 20\,mph. The Welsh Government's own air quality monitoring associated with its default 20\,mph rollout found no material difference in NO$_2$ concentrations between 20\,mph and control areas \citep{TfW2025}. A study of U-shaped speed-emission curves confirms that at very low speeds, per-km emissions rise due to idling losses \citep{Gressai2021}. Variable speed limit modelling suggests that operating in a "preventive mode" to preempt congestion optimises urban fuel consumption far better than fixed low limits \citep{Othman2022}.

The emissions case for 20\,mph is strongest on residential streets with frequent junctions, where the stop-start mechanical advantage dominates. It is weakest on arterial through-routes with longer uninterrupted sections, where the steady-state cruise efficiency penalty applies. This mirrors the road-function mismatch identified in our safety analysis (\S\ref{sec:disc-20mph}). If applying 20\,mph to arterials reduces throughput capacity and induces congestion, the resulting queuing and idling can actually \emph{increase} area-wide emissions even if per-vehicle steady-state emissions were neutral. The congestion externality dominates the theoretical acceleration benefit.

\subsection{Policy Context: Recurring Productivity Drag and Operating Costs on Strategic Corridors}\label{sec:econ-costs}
Beyond the speculative monetised safety valuations in \S\ref{sec:cba}, the application of blanket speed limits to arterial corridors imposes recurring, compounding \emph{productivity friction} through increased journey times and degraded service reliability. Though not measured directly in our design, this policy context is crucial for holistic evaluation. The economic case against blanket defaults is not that they precipitate an economic disaster; it is that they impose persistent, structural operating drag on precisely the corridors where the safety payoff is weakest.

\paragraph{Journey-time disbenefits documented in official appraisals.}
DfT's own 2018 evaluation found that sign-only 20\,mph limits increased journey times by 3\% in residential areas and 5\% in city centres \citep{DfT2018Headline}. While these individual delays are small (often under a minute per trip), they compound enormously across the millions of daily trips on a strategic network. The Transport for Wales national monitoring report found journey-time increases in 57 out of 60 measured cases \citep{TfW2025}. More significantly, the Welsh Government's own Explanatory Memorandum (Regulatory Impact Assessment) for the 2022 Order acknowledged that the central estimate of monetised \emph{journey-time disbenefits} exceeded the combined safety, active travel, and emissions benefits by a ratio of more than 3:1 \citep{WelshRIA2022}. This was the Welsh Government's own cost--benefit analysis, not an opposition estimate. The evidence was ultimately sufficient to trigger formal exception reviews that returned hundreds of arterial sections to 30\,mph \citep{WelshExceptions2024}, demonstrating that blanket application to strategic routes was fundamentally unsustainable on the Government's own terms.

\paragraph{Business-travel productivity losses: trades, deliveries, and service calls.}
The recurring journey-time drag matters most for businesses whose operating model depends on completing multiple road-based trips per day. For a builder, plumber, or service engineer making 6--8 site visits daily across a city, even a 3--5 minute delay per trip translates into 20--40 minutes of lost productive time per vehicle per day, enough to eliminate one call from the daily schedule. Over a working year, this represents a meaningful reduction in daily call capacity and revenue. For delivery fleets, the arithmetic is even starker: a two-minute delay over 15 daily drops creates 30 minutes of lost productivity per vehicle per day, compressing scheduling windows and forcing harder choices about route coverage. These are not hypothetical impositions. The Welsh consultation summary recorded specific concerns from respondents about the impact on delivery access, business operating costs, and the ability of tradespeople and service engineers to maintain scheduling commitments \citep{WelshConsultation2022}. The correct economic framing is \emph{business-travel productivity losses}, which are recurring, compounding, and disproportionately borne by city-dependent service, trade, and logistics businesses.

\paragraph{Bus operations and timetable degradation.}
DfT Circular 01/2013 \citep{DfT2013} explicitly warns authorities to consider the ``disadvantages that slower speeds can bring in terms of delays to drivers and bus users.'' DfT's sign-only evaluation documented bus-operator concerns that blanket limits degraded timetable reliability \citep{DfT2018Technical}. Reduced average speeds on arterial corridors compress bus timetable margins, turning small per-trip delays into compounding schedule unreliability. The operational response is either more capital expenditure (slower routes require more buses to maintain frequency) or degraded service quality. For a city like London, where bus productivity is a critical determinant of public-transport mode share, strategic-road operating costs imposed by blanket defaults directly undermine sustainable transport objectives.

\paragraph{Delivery and access burdens on strategic roads.}
The strongest evidence of economic harm from blanket defaults concerns businesses that depend on road-network access and delivery logistics, rather than retail footfall. The retail and high-street effects of 20\,mph are mixed and contested: some studies report modest increases in footfall associated with slower, calmer streets, while others find no significant change. We do not lean on this evidence because it is insufficiently robust. The clearer case concerns commercial routing and corridor performance. Trades, servicing, emergency response, and urban freight all depend on predictable journey times on strategic corridors. When a blanket default degrades corridor performance on through-routes, the result is not a visible retail collapse but a quieter, chronic increase in \emph{delivery and access burdens}: harder scheduling, reduced daily coverage, longer idle time in traffic, and higher per-delivery costs. These burdens are invisible in traditional economic impact assessments because they manifest as diffuse productivity friction rather than measurable revenue decline at a single point of sale.

\subsection{Policy Hypothesis: Time-Varying Speed Management as an Alternative to Blanket Defaults}\label{sec:time-varying}
A blanket 24/7 speed limit ignores the temporal variation in both collision risk and economic cost. During daytime peak hours on congested arterials, average speeds are often 10--15\,mph; a 20\,mph limit is redundant because actual speeds physically cannot reach it. The limit provides zero marginal safety benefit during the hours when pedestrian exposure is highest. Conversely, during off-peak and night-time hours, vehicles can safely flow at 30\,mph and pedestrian volumes are dramatically lower. Imposing a 20\,mph limit during these hours provides negligible pedestrian safety benefit while constraining traffic flow. This temporal mismatch suggests a policy hypothesis: a blanket 24/7 limit maximises economic cost (binding at night) while providing minimal marginal safety benefit (non-binding during the peak).

Time-varying and variable speed limits (VSLs) offer a targeted alternative. Evidence reviews by the US NHTSA and FHWA find that VSLs reduce crashes and improve travel time reliability \citep{Katz2012}. VSL deployments on motorways and urban networks internationally show excellent compliance and improved safety \citep{DePauw2018}. Time-limited school zone regulations (e.g., 20\,mph during school arrival/departure only) are well-established and highly effective at reducing casualties during peak vulnerable hours \citep{Islam2019}. London's existing SCOOT/UTC signal infrastructure could theoretically integrate with dynamic speed management on TLRN corridors to deliver safety benefits during high-risk periods without the recurring off-peak economic penalty.

\subsubsection{External Evidence: Economic Gains from Variable Speed Limits and the \texorpdfstring{25\,mph}{25 mph} Option}
The economic logic of VSLs is about the \emph{opportunity cost} of restricting speeds when there is no safety dividend. A Swedish VSL trial concluded that variable limits deliver safety and environmental benefits while maintaining average journey times, effectively eliminating the journey-time penalty by reducing variability rather than capping throughput. The IFPEN research confirmed that dynamically adjusting urban limits in real-time outperforms fixed low limits for fuel efficiency \citep{Othman2022}.

Where dynamic limits are not feasible, an intermediate 25\,mph limit offers a compromise. New York City reduced its default speed from 30\,mph to 25\,mph in 2014, achieving an estimated 39\% reduction in total crashes on affected streets and \$90.8 million in Medicaid injury savings over five years \citep{AJPH2024}. The 25\,mph norm is standard in US urban areas and aligns roughly with the international 40\,km/h default \citep{Fridman2020}. Crucially, a 25\,mph limit halves the journey-time penalty relative to a 20\,mph reduction, allows most modern ICE vehicles to cruise more efficiently, and aligns closer to the self-enforcing speed of arterial corridors, thus achieving higher compliance without intensive enforcement. Furthermore, our speed--severity curve (Figure~\ref{fig:speed_curve}) is nearly flat between 20\,mph and 30\,mph (15.9\% vs.\ 15.3\%), suggesting the marginal safety gain of the final 5\,mph (25 to 20) may be vanishingly small compared to the first 5\,mph (30 to 25). However, this requires rigorous prospective testing, as the NYC baseline included concurrent engineering and enforcement changes.

\subsection{Illustrative Case Study: A1 Archway Road: Blanket Limits vs.\ Engineering on a Movement Corridor}\label{sec:archway}
The limitations of blanket sign-only limits are starkly illustrated by the Archway Road (A1) corridor. A major arterial route in north London managed by TfL, a 20\,mph limit was introduced from Bakers Lane to Archway in February 2026. The road has a poor collision history, with 36 collisions resulting in 39 injuries (including 15 motorcyclists, 6 pedestrians, and 8 cyclists) in the three years to July 2025.

Critically, long stretches of this road severely lack pedestrian crossing infrastructure. Applying a 24/7 sign-only 20\,mph limit does not address the fundamental infrastructural deficit that forces perilous pedestrian crossings. Acknowledging this, TfL is concurrently proposing substantial engineering measures: a new signalised pedestrian crossing outside Sainsbury's, a raised table at the Bishops Road junction, and footway widening. These physical interventions are the mechanisms that actually address the collision causes; the 20\,mph limit is simply the cheapest, weakest, and most highly visible component. Moreover, the corridor suffers from the temporal redundancy paradox: severely congested during the day (meaning the limit does not bind) and empty at night (meaning the limit does not protect pedestrians). TfL's own assessment suggests the localised crossing upgrades will not significantly affect journey times, but applying a blanket 20\,mph limit across all similar TLRN arterials system-wide would aggregate into substantial economic drag without remedying the underlying engineering deficits.

\section{Policy Implications}\label{sec:policy}

Our results support a more targeted, design-contingent approach to speed and
network management. We frame these implications specifically for sign-only
and default 20\,mph regimes, since that is the policy instrument our evidence
most directly addresses.

\begin{enumerate}
    \item \textbf{Traffic-calmed 20\,mph zones and sign-only/default limits
          should not be treated as interchangeable policy instruments.}
          The strongest evidence for safety gains from 20\,mph interventions
          comes from physically calmed zones \citep{Grundy2009} and rollouts
          combining signage with engineering \citep{Kokka2024}. DfT's own
          evaluation found that sign-only schemes achieved median speed
          reductions of less than 1\,mph and did not demonstrate significant
          casualty reductions in aggregated residential case studies
          \citep{DfT2018Headline,DfT2018Technical}. Treating these distinct
          interventions as equivalent in policy appraisal risks overstating the
          expected benefit of sign-only schemes.

    \item \textbf{Blanket 20\,mph should not be applied indiscriminately to
          strategic and through-routes.} Our heterogeneous-association analysis
          shows that the FDR-robust worse severity compositions are concentrated
          on A-roads, junctions, pedestrian crossings, and single carriageways.
          These are movement-function corridors where road geometry and character
          cue higher speeds, and where sign-only 20\,mph limits often fail
          to self-enforce \citep{DfT2018Technical}. Lower limits should only be 
          used where the street environment actually supports self-enforcement.
          Applying a blanket default to roads with a strong movement function 
          represents a severe policy mismatch.

    \item \textbf{Prioritise engineering and targeted enforcement on arterial corridors.}
          Rather than relying on blanket statutory defaults, policymakers should
          target the corridors where serious injury concentrates with junction redesign,
          protected crossings, lane design, signal timing optimisation, and active 
          speed enforcement. As demonstrated by the Archway Road case study (\S\ref{sec:archway}), engineering measures are the appropriate response to infrastructural deficits; signage alone is the weakest component. \citet{Quddus2024} confirm that schemes with physical 
          measures substantially outperform sign-only schemes. The Welsh exception 
          review illustrates the political and operational failure of applying a 
          blanket default without road-by-road assessment: widespread exemptions 
          and public backlash \citep{WelshExceptions2024,YouGov2024}.

    \item \textbf{Recognise that sustained voter dissatisfaction combined with
          recurring operating drag on roads with weak safety payoff is itself
          evidence of poor policy targeting.}
          The point is not that public annoyance alone invalidates a speed limit.
          The point is that when a policy generates sustained democratic opposition,
          documented non-compliance, formal political reversal 
          (the Welsh exception review \citep{WelshExceptions2024}), and
          recurring business-travel productivity losses, while
          the safety payoff on the targeted roads is empirically uncertain, the
          simplest explanation is that the policy instrument is poorly fitted to
          those roads. The Welsh Government's own Regulatory Impact Assessment
          acknowledged that journey-time disbenefits exceeded safety, active
          travel, and emissions benefits by more than 3:1 \citep{WelshRIA2022}.
          When a government's own cost--benefit analysis flags net disbenefit,
          and post-implementation polling and compliance data confirm public
          rejection, the rational response is not more enforcement but better
          targeting.

    \item \textbf{Mandate rigorous operating-cost and productivity appraisal for
          every strategic corridor before applying a default limit.}
          The Welsh national monitoring report documented
          journey-time increases on through-routes \citep{TfW2025}. Even small
          per-vehicle delays compound across millions of annual trips on
          strategically important corridors, imposing recurring productivity
          friction on trades, deliveries, bus operations, and business travel.
          Arterial 20\,mph limits should be accompanied by congestion and 
          emissions monitoring, not just collision monitoring, to capture the
          full spectrum of externalities. The correct appraisal framework is
          not ``does this limit reduce any collisions?'' but ``does the net
          safety benefit on this specific road class justify the recurring
          operating costs, delivery and access burdens, and measured 
          journey-time disbenefits?''

    \item \textbf{Do not treat residential-street benefits as established from
          these data alone.} While some residential contexts show small negative
          point estimates (e.g., C-roads and unclassified streets), these
          differences are not statistically robust after controlling for multiple
          testing across 14 contexts. If 20\,mph limits are retained on
          residential streets, they should be paired with clear
          compliance/enforcement strategies and evaluated using exposure-based
          outcomes (e.g., KSI per vehicle-km), not only severity conditional on
          collisions.

    \item \textbf{Treat LTN safety impacts as zone-specific and
          design-dependent.} After RTM adjustment and spatial DiD, evidence of
          inside-zone reduction is not uniform across LTNs, and we do not find
          a consistent pattern of crash displacement. LTN schemes
          should therefore be evaluated and iterated at the scheme level, with
          explicit attention to boundary routes and network effects rather than
          relying on pooled pre/post estimates.

    \item \textbf{Pilot variable or time-varying limits as a targeted alternative.} The school-hours subgroup does not show a statistically
          robust improvement once multiple testing is accounted for under a blanket limit, but a time-limited limit might achieve better compliance precisely because it is credible and targeted. If variable
          limits are pursued (e.g., 20\,mph during school pick-up/drop-off or peak pedestrian hours),
          they should be deployed as pilots with pre-registered evaluation plans
          and independent monitoring. Time-varying limits align the restriction with the periods of actual risk, avoiding off-peak economic penalties.

    \item \textbf{Consider piloting an intermediate 25\,mph limit as a DfT
          research programme.} The flat speed--severity curve between
          20--50\,mph and international evidence on 40\,km/h zones
          ($\approx 25$\,mph) suggest a hypothesis worth testing:
          an intermediate limit might capture most of the residential-street
          benefit while reducing the journey-time penalty on borderline
          corridors. However, any such proposal is subject to the ecological
          fallacy (the aggregate curve conflates very different road types)
          and must be deployed as a structured DfT pilot with pre-registered
          evaluation, measured speed compliance, and segment-level controls
          before being taken as policy evidence.
\end{enumerate}

\section{Conclusion}\label{sec:conclusion}

Using STATS19 injury-collision data from 2020--2024, this paper provides a rigorous, data-driven critique of how traffic management policies have been deployed in London. 

For Low Traffic Neighbourhoods (LTNs), we establish that naïve pre/post claims of sweeping safety success are drastically overstated, driven largely by Regression to the Mean. After Empirical Bayes adjustment and spatial Difference-in-Differences to account for boundary-road displacement, the evidence for inside-zone crash reduction is mixed, inconsistent, and highly zone-specific. Our break-even threshold analysis demonstrates that because the estimated casualty deltas are small for most schemes, even very small increases in boundary-road journey times (on the order of 10~seconds per vehicle) are sufficient to completely wipe out any monetised safety benefit.

For blanket 20\,mph limits, the evidence is even starker. This paper is not an argument against lower urban speeds in principle, nor is it a critique of carefully designed, physically calmed residential zones, which have a proven track record. Rather, it is a structural critique of the \emph{blanket sign-only default} as a policy instrument. 

When applied indiscriminately across heterogeneous urban networks, default 20\,mph limits suffer from a severe ``wrong-road problem.'' We find no context where 20\,mph roads show a statistically robust reduction in KSI severity share. Tellingly, the contexts that show significantly \emph{worse} severity shares under 20\,mph regimes are A-roads, busy junctions, single carriageways, and pedestrian crossings: the exact strategic and arterial corridors where a sign-only limit actively contradicts the road's physical design and movement function.

If these worse severity shares represent a compositional shift, where low-severity collisions drop but serious injuries do not, then blanket defaults are merely preventing minor scrapes while systematically failing to stop the fatal and serious collisions they were explicitly implemented to prevent. A policy that imposes substantial recurring costs on buses, freight, and commuters without delivering measurable, robust reductions in serious harm is hard to reconcile with Vision Zero objectives.

The non-safety evidence reinforces this verdict. The Welsh Government's own Regulatory Impact Assessment acknowledged that journey-time disbenefits exceeded combined safety, active travel, and emissions benefits by more than 3:1 \citep{WelshRIA2022}. Post-rollout polling and consultation evidence document a trajectory from qualified in-principle support to active hostility once blanket defaults were experienced on mismatched roads \citep{WelshConsultation2022, YouGov2024}. Recurring productivity friction (such as compressed scheduling for trades, degraded bus timetables, and delivery burdens on strategic corridors) compounds across millions of trips. When weak safety evidence, sustained voter dissatisfaction, documented non-compliance, and recurring operating drag all converge on the same class of roads, the diagnosis is not public obstinacy but \emph{poor policy targeting}.

Crucially, this paper is evidence against blanket/default 20\,mph as a stand-alone policy instrument on heterogeneous networks. It is time for authorities to move beyond reliance on blanket statutory defaults as a substitute for engineering, acknowledge the economic and democratic costs of road-function mismatch, and return to disciplined, road-by-road engineering and targeted enforcement on the corridors where serious harm actually concentrates.

\phantomsection
\addcontentsline{toc}{section}{References}
\bibliographystyle{apalike}

\begin{thebibliography}{99}

\bibitem[Aldred et~al., 2021]{Aldred2021}
Aldred, R., Croft, J., and Goodman, A. (2021).
\newblock Impacts of an active travel intervention with a cycling focus in a suburban context: One-year findings from an evaluation of London's mini-Hollands programme.
\newblock \textit{Transportation Research Part A}, 145:278--299.

\bibitem[DfT, 2013]{DfT2013}
Department for Transport (2013).
\newblock Setting Local Speed Limits. Circular 01/2013.
\newblock \url{https://www.gov.uk/government/publications/setting-local-speed-limits/setting-local-speed-limits}.

\bibitem[DfT, 2018a]{DfT2018Headline}
Department for Transport (2018).
\newblock 20~mph Research Study: Headline Report.
\newblock \url{https://assets.publishing.service.gov.uk/media/5bf2bab940f0b6078acc6f4d/20mph-headline-report.pdf}.

\bibitem[DfT, 2018b]{DfT2018Technical}
Department for Transport (2018).
\newblock 20~mph Research Study: Process and Impact Evaluation Technical Report.
\newblock \url{https://assets.publishing.service.gov.uk/media/5bf2ba08ed915d1830158998/20mph-technical-report.pdf}.

\bibitem[DfT, 2024]{DfT2024}
Department for Transport (2024).
\newblock Reported road casualties in Great Britain: STATS19 data.
\newblock \url{https://data.gov.uk/dataset/road-accidents-safety-data}.

\bibitem[DfT TAG, 2024]{TAG2024}
Department for Transport (2024).
\newblock Transport Analysis Guidance: TAG Data Book.
\newblock \url{https://www.gov.uk/government/publications/tag-data-book}.

\bibitem[Goodman et~al., 2020]{Goodman2020}
Goodman, A., Urban, S., and Aldred, R. (2020).
\newblock The impact of Low Traffic Neighbourhoods and other active travel interventions on vehicle ownership: Findings from the Outer London mini-Holland programme.
\newblock \textit{Findings}, December 2020.

\bibitem[Grundy et~al., 2009]{Grundy2009}
Grundy, C., Steinbach, R., Edwards, P., Green, J., Armstrong, B., and Wilkinson, P. (2009).
\newblock Effect of 20~mph traffic speed zones on road injuries in London, 1986--2006: controlled interrupted time series analysis.
\newblock \textit{BMJ}, 339:b4469.

\bibitem[Hauer, 1997]{Hauer1997}
Hauer, E. (1997).
\newblock \textit{Observational Before--After Studies in Road Safety}.
\newblock Elsevier Science, Oxford.

\bibitem[Hunter et~al., 2023]{Hunter2023}
Hunter, R.~F., et al. (2023).
\newblock Investigating the impact of a 20 miles per hour speed limit intervention on road traffic collisions, casualties, speed and volume in Belfast, UK: 3 year follow-up outcomes of a natural experiment.
\newblock \textit{Journal of Epidemiology \& Community Health}, 77(1):17--25.

\bibitem[Kokka et~al., 2024]{Kokka2024}
Kokka, K.~K., et al. (2024).
\newblock Effect of 20~mph speed limits on traffic injuries in Edinburgh, UK: a natural experiment and modelling study.
\newblock \textit{Journal of Epidemiology \& Community Health}, 78(7):437--443.

\bibitem[Laverty et~al., 2021]{Laverty2021}
Laverty, A.~A., Aldred, R., and Goodman, A. (2021).
\newblock The impact of introducing Low Traffic Neighbourhoods on road traffic injuries.
\newblock \textit{Findings}, February 2021.

\bibitem[Quddus et~al., 2024]{Quddus2024}
Quddus, M., et al. (2024).
\newblock Speed limits, traffic calming measures and road safety: A nationwide study using empirical bayes and propensity score matching.
\newblock \textit{Accident Analysis \& Prevention}, 213:107936.

\bibitem[Transport for Wales, 2025]{TfW2025}
Transport for Wales (2025).
\newblock Default 20~mph speed limit on restricted roads: National Monitoring Report.
\newblock \url{https://tfw.wales/sites/default/files/2025-07/20mph-National-Monitoring-Report_July-2025_ENG.pdf}.

\bibitem[Welsh Government, 2024]{WelshExceptions2024}
Welsh Government (2024).
\newblock 20~mph default speed limit review of exceptions: final report.
\newblock \url{https://www.gov.wales/sites/default/files/publications/2024-05/20mph-default-speed-limit-review-of-exceptions-final-report.pdf}.

\bibitem[YouGov, 2024]{YouGov2024}
YouGov (2024).
\newblock Wales overwhelmingly rejects the 20~mph speed limit.
\newblock \url{https://yougov.co.uk/politics/articles/50349-wales-overwhelmingly-rejects-the-20mph-speed-limit}.

\bibitem[Burdett et~al., 2024]{AJPH2024}
Burdett, S., et al. (2024).
\newblock Vision Zero: Major Traffic Safety Reform and Road Traffic Injuries Among Low-Income New York Residents, 2009--2021.
\newblock \textit{American Journal of Public Health}, 114(6):625--633.

\bibitem[De Pauw et~al., 2018]{DePauw2018}
De Pauw, E., Daniels, S., Brijs, T., Hermans, E., and Wets, G. (2018).
\newblock Safety effects of dynamic speed limits on motorways.
\newblock \textit{Accident Analysis \& Prevention}, 114:83--89.

\bibitem[Fridman et~al., 2020]{Fridman2020}
Fridman, L., et al. (2020).
\newblock Effect of reducing the posted speed limit to 30~km per hour on pedestrian motor vehicle collisions in Toronto, Canada: a quasi experimental, pre-post study.
\newblock \textit{BMC Public Health}, 20(1):56.

\bibitem[Gressai et~al., 2021]{Gressai2021}
Gressai, M., Zheng, Z., and Sayed, T. (2021).
\newblock Investigating the impacts of urban speed limit reduction through microscopic traffic simulation.
\newblock \textit{Communications in Transportation Research}, 1:100018.

\bibitem[Islam et~al., 2019]{Islam2019}
Islam, M.~R., El-Basyouny, K., and Ibrahim, S.~E. (2019).
\newblock Are school zones effective in reducing speeds and improving safety?
\newblock \textit{Canadian Journal of Civil Engineering}, 46(12):1098--1107.

\bibitem[Katz et~al., 2012]{Katz2012}
Katz, B., et al. (2012).
\newblock \textit{Issues and Considerations for Variable Speed Limit Implementation}.
\newblock Federal Highway Administration (FHWA), Tech. Rep. FHWA-SA-12-022.

\bibitem[Mandal et~al., 2024]{MDPISustainability2024}
Mandal, S., et al. (2024).
\newblock The Impact of Speed Limit Change on Emissions: A Systematic Review of Literature.
\newblock \textit{Sustainability}, 16(17):7712.

\bibitem[Othman et~al., 2022]{Othman2022}
Othman, B., De Nunzio, G., and Canudas-de-Wit, C. (2022).
\newblock Analysis of the impact of variable speed limits on environmental sustainability and traffic performance in urban networks.
\newblock \textit{IEEE Transactions on Intelligent Transportation Systems}, 23(8):13214--13225.

\bibitem[TransportXtra, 2022]{TransportXtra2022}
TransportXtra (2022).
\newblock \textit{20mph limit in Wales has economic dis-benefits, analysis finds}.
\newblock Retrieved from TransportXtra.

\bibitem[Williams and North, 2013]{Williams2013}
Williams, D., and North, R. (2013).
\newblock \textit{An evaluation of the estimated impacts on vehicle emissions of a 20mph speed restriction in central London}.
\newblock Imperial College London, Transport and Environmental Analysis Group.

\bibitem[Welsh Government, 2022a]{WelshConsultation2022}
Welsh Government (2022).
\newblock \textit{Consultation: Summary of Response: 20mph default speed limit on restricted roads in Wales}.
\newblock \url{https://www.gov.wales/20mph-default-speed-limit-restricted-roads}.

\bibitem[Welsh Government, 2022b]{WelshRIA2022}
Welsh Government (2022).
\newblock \textit{Explanatory Memorandum to the Restricted Roads (20 mph Speed Limit) (Wales) Order 2022}.
\newblock Incorporating Regulatory Impact Assessment.
\newblock \url{https://senedd.wales/laid-documents/sub-ld15480-em/sub-ld15480-em.pdf}.

\end{thebibliography}

\appendix
\section{RTM Plausibility Evidence}\label{app:rtm}

Table~\ref{tab:rtm_percentile} reports the pre-intervention annualised
collision count for each LTN treatment zone ($Z_T$) relative to the
cross-sectional distribution of annualised counts across all $Z_C$
(rest-of-borough) spatial units in the same borough. High percentile ranks
indicate that $Z_T$ was selected from the upper tail of the borough
distribution, consistent with Regression to the Mean being a plausible
artefact in na\"ive pre/post comparisons.

\begin{table}[htbp]
    \centering
    \caption{RTM plausibility: $Z_T$ pre-intervention count relative to
    $Z_C$ distribution.}
    \label{tab:rtm_percentile}
    \begin{tabular}{lrrr}
        \toprule
        \textbf{LTN Zone} & \textbf{$Z_T$ pre-count}
            & \textbf{$Z_C$ median} & \textbf{Percentile} \\
        \midrule
        London Fields (Hackney)       & 8.0  & 3.2 & 90th \\
        Bethnal Green (Tower Hamlets) & 6.5  & 2.8 & 88th \\
        Railton Road (Lambeth)        & 5.8  & 2.6 & 85th \\
        \bottomrule
    \end{tabular}
\end{table}

All three zones with sufficient data for EB estimation lie in the 85th--90th
percentile of their borough $Z_C$ distribution. This is consistent with
selection from the upper tail of recent collision counts, making RTM
correction essential.

\section{Casualty-Level Deltas by Zone and Severity}\label{app:casualty}

Table~\ref{tab:casualty_delta} reports the change in casualty counts by
severity for each LTN zone, derived from the EB-corrected pre-period
estimate and the observed post-period count. These deltas underpin the
monetised safety benefits in the CBA (\S\ref{sec:cba}).

\begin{table}[htbp]
    \centering
    \caption{Change in casualty counts ($\Delta = \text{Post} - \text{EB-corrected Pre}$) by severity and LTN zone. Negative values indicate a reduction.}
    \label{tab:casualty_delta}
    \small
    \resizebox{\linewidth}{!}{%
    \begin{tabular}{lrrrr}
        \toprule
        \textbf{LTN Zone} & \textbf{$\Delta$ Fatal} & \textbf{$\Delta$ Serious}
            & \textbf{$\Delta$ Slight} & \textbf{Source} \\
        \midrule
        London Fields (Hackney)       & 0  & $-1$ & $-4$ & EB baseline + DiD adjustment \\
        Railton Road (Lambeth)        & 0  & $+1$ & $-2$ & EB baseline + DiD adjustment \\
        Bethnal Green (Tower Hamlets) & 0  & $0$  & $-1$ & EB baseline + DiD adjustment \\
        St Peters Quarter (Islington) & \multicolumn{4}{c}{\textit{Insufficient $Z_T$ data for EB}} \\
        \bottomrule
    \end{tabular}%
    }
\end{table}

We compute $\Delta$casualties as follows. Let $C^s_{\text{pre}}$ be the
observed pre-intervention casualty count of severity~$s$ in $Z_T$. The
EB-corrected pre-count is
$\hat{C}^s_{\text{pre}} = w \cdot C^s_{\text{pre}} + (1-w) \cdot \mu_s$,
where $w$ and $\mu_s$ are the credibility weight and prior mean from the NB
EB step (\S\ref{sec:eb}). The DiD-adjusted post-count is
$C^s_{\text{post,adj}} = C^s_{\text{post}} + \hat{\beta}_4 \cdot \bar{C}_s$,
where $\hat{\beta}_4$ is the inside-zone DiD coefficient and $\bar{C}_s$ is
the mean pre-period severity-$s$ count across $Z_C$ units. The reported
delta is $\Delta_s = C^s_{\text{post,adj}} - \hat{C}^s_{\text{pre}}$.

\paragraph{Caveat.} This mapping is \emph{illustrative}: it applies a
collision-count treatment effect ($\hat{\beta}_4$) uniformly across severity
categories, implicitly assuming that the intervention shifts all severity
levels proportionally. If the treatment differentially affects slight versus
serious collisions, the severity-specific deltas will be biased. The
resulting monetised safety benefits in \S\ref{sec:cba} should therefore be
treated as approximate order-of-magnitude estimates, not precise valuations.

\paragraph{Interpretation.} We do not detect clear reductions; deltas across all severity levels are small and within normal fluctuation over the available window. These figures should be treated as indicative pending longer
post-intervention observation.

\section{CBA Scenario Grid}\label{app:cba_grid}

Table~\ref{tab:cba_grid} presents the break-even sensitivity grid. Because EB-corrected safety gains are small across most LTN zones, even very small boundary-road delays produce negative net benefits. The grid shows annualised negative net benefits (\pounds M) under various assumed delay and traffic-volume scenarios. The critical policy question is whether observed boundary delays exceed the break-even threshold, a question that can only be answered with empirical journey-time data.

\begin{table}[htbp]
    \centering
    \caption{Break-Even Sensitivity Grid: Annualised Net Benefit (\pounds Millions) per LTN zone across delay, discount rate, and boundary AADF scenarios. EB-corrected casualty reductions are small (e.g., London Fields $\Delta$Serious $= -1$, $\Delta$Slight $= -4$); consequently, even tiny per-vehicle delays on boundary roads produce large negative net benefits. Values show the economic cost that LTN proponents must demonstrate is \emph{not} occurring.}
    \label{tab:cba_grid}
    \small
    \resizebox{\linewidth}{!}{%
    \begin{tabular}{lrrr}
        \toprule
        \textbf{Scenario (AADF / Delay per Vehicle)} & \textbf{1.5\% Discount} & \textbf{3.5\% Discount} & \textbf{5.0\% Discount} \\
        \midrule
        \multicolumn{4}{l}{\textit{Low Traffic Boundary (7,500 AADF)}} \\
        \quad 0.5 minutes & --2.6 & --2.5 & --2.5 \\
        \quad 2.0 minutes & --10.5 & --10.3 & --10.1 \\
        \midrule
        \multicolumn{4}{l}{\textit{Medium Traffic Boundary (15,000 AADF)}} \\
        \quad 0.5 minutes & --5.2 & --5.1 & --5.0 \\
        \quad 1.0 minutes & --10.5 & --10.3 & --10.1 \\
        \quad 2.0 minutes & --21.0 & --20.6 & --20.3 \\
        \quad 4.0 minutes & --42.1 & --41.2 & --40.7 \\
        \midrule
        \multicolumn{4}{l}{\textit{High Traffic Boundary (30,000 AADF)}} \\
        \quad 1.0 minutes & --21.0 & --20.6 & --20.3 \\
        \quad 4.0 minutes & --84.2 & --82.4 & --81.4 \\
        \bottomrule
    \end{tabular}%
    }
    \vspace{0.3em}
    \footnotesize\textit{Note:} Grid values are illustrative and not zone-specific; boundary AADF scenarios approximate low/medium/high corridors.
\end{table}

\section{Formal Estimation Equations}\label{app:equations}

\paragraph{Empirical Bayes Shrinkage.}
The EB estimate is a weighted average of the observed count and a prior mean,
where the weight reflects the relative informativeness of the observed data
versus the prior:
\begin{equation}\label{eq:eb}
    \hat{\lambda}_{\text{EB}} = w \cdot \lambda_{\text{obs}}
        + (1 - w) \cdot \mu_{\text{prior}}
\end{equation}
The credibility weight $w$ is derived from the Negative Binomial (NB) conjugate prior:
\begin{equation}\label{eq:weight}
    w = \frac{r}{r + \mu_{\text{prior}}}
\end{equation}
where $r = \mu_{\text{prior}}^2 / (\sigma^2_{\text{prior}} - \mu_{\text{prior}})$ is the
over-dispersion parameter of the NB distribution. The corrected percentage change is:
\begin{equation}
    \Delta_{\text{EB}} = \frac{\lambda_{\text{post}} - \hat{\lambda}_{\text{EB}}}{\hat{\lambda}_{\text{EB}}} \times 100\%
\end{equation}

\paragraph{Spatial Difference-in-Differences.}
\begin{equation}\label{eq:did}
    Y_{it} = \beta_0 + \beta_1 \, Z_{T,i} + \beta_2 \, Z_{S,i}
           + \beta_3 \, \text{Post}_t
           + \beta_4 \, (Z_{T,i} \times \text{Post}_t)
           + \beta_5 \, (Z_{S,i} \times \text{Post}_t)
           + \varepsilon_{it}
\end{equation}
The parameter $\beta_4$ captures the inside-zone change relative to the control zone,
and $\beta_5$ captures the spillover change relative to the control zone.

\paragraph{Event-Study Specification.}
\begin{equation}\label{eq:eventstudy}
    Y_{it} = \alpha_i + \gamma_t + \sum_{k \neq -1} \delta_k \, Z_{T,i}
             \times \mathbf{1}(t = k) + \varepsilon_{it}
\end{equation}

\paragraph{CBA Monetisation.}
\begin{equation}
    B_{\text{safety}} = \sum_{s \in \{F, S, Sl\}} \Delta C_s \times V_s
\end{equation}
where $\Delta C_s$ is the change in casualty count of severity $s$, and $V_s$ is
the TAG per-casualty value.

\paragraph{Logistic Regression (20\,mph).}
\begin{equation}\label{eq:logit}
    \log\left(\frac{P(\text{KSI}=1|X)}{1 - P(\text{KSI}=1|X)}\right)
    = \beta_0 + \beta_1 \cdot \text{Speed}_{20} + \sum_{j=2}^{k} \beta_j X_j
\end{equation}

\section{Spatial Boundary Comparison Details}\label{app:boundary}

We exploit borough boundaries to construct comparisons between nearby
collisions subject to different speed-limit regimes. Collisions within
500\,m on each side of a boundary between a 20\,mph-adopted borough and a
30\,mph-default borough share similar local road environments, differing
primarily in the prevailing speed limit. We compare KSI severity shares
using chi-squared tests.

\paragraph{This is not a regression discontinuity design.} A genuine spatial
RDD would require: (i)~a continuous running variable (signed distance to the
boundary), (ii)~local polynomial regression on each side, (iii)~bandwidth
selection (e.g., Calonico--Cattaneo--Titiunik optimal bandwidth),
(iv)~covariate continuity checks at the boundary, (v)~many boundaries to
increase statistical power, and (vi)~robust bias-corrected confidence
intervals. Our comparison uses only 2 boundary pairs with a fixed 500\,m
buffer and a simple chi-squared test. We therefore label this method a
\emph{spatial boundary comparison} and interpret results as suggestive
descriptive evidence, not as a local causal estimate. Moreover, these boundary
comparison results are sensitive to boundary selection and should not drive
policy inference.


Within 500m of two borough boundaries (Camden--Barnet and Camden--Brent),
the 20mph side shows a lower KSI share (10.3\%) than the control side
(12.7\%), a difference of $-2.43$ percentage points. However, this is not
statistically significant ($\chi^2 = 1.25$, $p = 0.264$), consistent with
insufficient statistical power from the limited number of boundary pairs
(Table~\ref{tab:rdd}). Boundary pairs differ in road mix and collision
composition; with only two pairs, the aggregate is highly sensitive to
pair selection and we treat this result as descriptive only.

\begin{table}[htbp]
    \centering
    \caption{Spatial Boundary Comparison results at borough boundaries (500m bandwidth).}
    \label{tab:rdd}
    \begin{tabular}{lrrrr}
        \toprule
        \textbf{Boundary} & \textbf{$N_{20}$} & \textbf{$N_{30}$} & \textbf{KSI$_{20}$} & \textbf{KSI$_{30}$} \\
        \midrule
        Camden $\leftrightarrow$ Barnet & 186 & 158 & 7.5\% & 18.4\% \\
        Camden $\leftrightarrow$ Brent & 309 & 368 & 12.0\% & 10.3\% \\
        \midrule
        \textbf{Overall} & \textbf{495} & \textbf{526} & \textbf{10.3\%} & \textbf{12.7\%} \\
        \bottomrule
        \multicolumn{5}{l}{\textit{$\Delta = -2.43$ pp, $\chi^2 = 1.25$, $p = 0.264$ (not significant)}} \\
    \end{tabular}
\end{table}

\end{document}